\documentclass[aps,prb,twocolumn,superscriptaddress]{revtex4-1}
\usepackage{amsmath, graphicx, dcolumn, bm}
\usepackage[pdfencoding=auto, psdextra]{hyperref}
\usepackage[normalem]{ulem}
\usepackage[dvipsnames]{xcolor}
\usepackage{soul}
\definecolor{purp}{HTML}{832591}

\hypersetup{
  colorlinks,
  allcolors = {purp},
  citecolor = {purp},
  allbordercolors = {white},
}

\begin{document}

\title{Density-Jump Transitions in the Debye-H\"uckel Theory of Spin Ice and Electrolytes}

\author{Omar J. Abbas}
\affiliation{London Centre for Nanotechnology and Department of Physics and Astronomy, University College London, 17-19 Gordon Street, London WC1H 0AH, United Kingdom}
\affiliation{Current address: London Centre for Nanotechnology and Department of Electronics and Electronic Engineering, University College London, 17-19 Gordon Street, London WC1H 0AH, United Kingdom}

\author{Steven T. Bramwell}
\affiliation{London Centre for Nanotechnology and Department of Physics and Astronomy, University College London, 17-19 Gordon Street, London WC1H 0AH, United Kingdom}

\author{Daan M. Arroo}
\email{d.arroo14@imperial.ac.uk}
\affiliation{London Centre for Nanotechnology and Department of Physics and Astronomy, University College London, 17-19 Gordon Street, London WC1H 0AH, United Kingdom}
\affiliation{Department of Materials, Imperial College London, Exhibition Road, London SW7 2AZ, United Kingdom}

\begin{abstract}

Debye-H\"{u}ckel theory, originally developed to describe dilute electrolyte solutions, has proved particularly successful as a description of magnetic monopoles in spin ice systems such as Dy$_{2}$Ti$_{2}$O$_{7}$. For this model, Ryzhkin \emph{et al.} predicted a phase transition in which the monopole density abruptly changes by several orders of magnitude but to date this transition has not been observed experimentally. Here we confirm that this transition is a robust prediction of Debye-H\"{u}ckel theory, that does not rely on approximations made in the previous work. However, we also find that the transition occurs in a regime where the theory breaks downs as a description of a Coulomb fluid and may be plausibly interpreted as an indicator of monopole crystallisation. By extending Ryzhkin's model, we associate the density jump of Debye-H\"uckel theory with the monopole crystallisation observed in staggered-potential models of `magnetic moment fragmentation', as well as with crystallisation in conserved monopole-density models. The possibility of observing a true density-jump transition in real spin ice and electrolyte systems is discussed.

\end{abstract}
\maketitle

\section{Introduction}

Frustrated magnetic materials have long been a focus of interest as systems in which large ground-state degeneracies can lead to the appearance of exotic states \cite{Ramirez1994}. Prominent among these are spin ice systems \cite{CMSRev,Bramwell2020} such as Dy$_{2}$Ti$_{2}$O$_{7}$ (DTO) and Ho$_{2}$Ti$_{2}$O$_{7}$ (HTO), in which dipolar and exchange interactions between Ising-like rare-earth ions on a pyrochlore lattice lead to an extensively degenerate low temperature state in which each vertex of the pyrochlore lattice has two spins pointing in and two pointing out. This condition maps to the ``ice rules'' that govern proton disorder in water ice and hence spin ice and water ice share the same characteristic residual entropy per site first estimated by Pauling \cite{Pauling1935} in 1935:
\begin{equation}
    s_0 \approx k_{\rm B} \ln \left(\frac{3}{2}\right).
\end{equation}

While at a basic level the natural models with which to describe such systems are the vertex models \cite{Baxter} that were inspired by this feature of water ice \cite{Lieb1967}, a key insight in the study of spin ice systems has been that they can be usefully described in terms of an emergent Coulomb phase \cite{Henley2010} where spins are identified with the flux of a divergenceless field and ice rule defects carry a magnetic charge. Statics and dynamics can then be represented in terms of free magnetic monopoles that interact through a magnetic Coulomb interaction \cite{Ryzhkin2005,CMS2008,Castelnovo2021}.

Taking this picture of interacting magnetic charges seriously, the thermodynamics of monopoles in spin ice is elegantly captured by a ``magnetolyte'' model \cite{CMS2011} whose properties can be analysed in terms of Debye-H\"uckel theory, as applied to weak electrolytes \cite{DebyeHuckel1923,Moore1964}. The excitation of singly- and doubly-charged monopoles (3:1 vertices and all-in/all-out vertices, respectively) in the grand canonical ensemble is then analogous to ionisation in an electrochemical system of the form
\begin{equation}
    \rm 2H_{2}O \rightleftharpoons H_{3}O^{+} + OH^{-} \rightleftharpoons H_{4}O^{2+} + O^{2-}
\end{equation}
with the density of monopoles controlled by their respective chemical potentials. An equilibrium is reached in which charge correlations lead to an exponential screening of the Coulomb interactions between monopoles. When the grand canonical vacuum is identified as the Pauling ice state (with the associated residual entropy), Debye-H\"{u}ckel theory applied to spin ice in this way provides remarkably good agreement with experiment \cite{Kaiser2018} and the theory has been widely used to describe a broad range of spin ice systems \cite{Zhou2011,Kirschner2018,Farhan2019}.

There nevertheless remains an unresolved point of tension between the predictions of the Debye-H\"{u}ckel magnetolyte model of spin ice and experimental measurements of real spin ice systems. Using parameters for HTO, an early work by Ryzhkin \emph{et al.}~\cite{Ryzhkin2012} shows that Debye-H\"{u}ckel theory predicts a first-order phase transition at $T_{\rm m} \approx 0.1887$ K in which the monopole density abruptly jumps by several orders of magnitude. A similar transition had previously been predicted within the framework of Debye-H\"{u}ckel theory by Kozlov \emph{et al.} \cite{kozlov1990phase}. Since Debye-H\"{u}ckel theory is generally in excellent agreement with experiment as a model of spin ice systems, it is surprising that no experimental signatures of such a transition have been observed despite numerous studies having probed the relevant temperature regime for HTO \cite{Harris1997,Matsuhira2000,Clancy2009,Paulsen2019}. It may be relevant that Ryzhkin \emph{et al.} use certain approximations to the Debye-H\"uckel free energy and one might suspect that these either introduced a transition that does not occur without these approximations, or shifted it from a parameter range that has not yet been accessed experimentally. Hence it is useful to re-examine the problem using the Debye-H\"uckel theory developed by Kaiser \emph{et al}. ~\cite{Kaiser2018}, recapped below in Section II, which dispenses with these approximations.

The first result of this paper, described in section III, is that the approximations made in the previous work do not introduce the transition, but removing them shifts it into a parameter range that is far from that applicable to HTO. More importantly, the transition occurs in a range where Debye-H\"uckel theory is no longer a formally valid description of the lattice Coulomb fluid, and where instead, there is monopole crystallisation~\cite{BrooksBartlett2014,Raban2019}.
We observe, however (section IV), that the Debye-H\"uckel theory nevertheless retains merit as a qualitative description of the lattice Coulomb fluid in this range, capturing important properties described in Ref. \cite{Raban2019} and clarifying how the model in this reference relates to an alternative model of crystallisation proposed by Borzi \emph{et al.} \cite{Borzi2013}  The general conclusion (section V), then, is that either the transition of Ref. \cite{Ryzhkin2012} is a qualitative analogue of the crystallisation, or else it is a genuine liquid-liquid transition, but one that will be masked by crystallisation in a real lattice Coulomb fluid. We conclude by briefly speculating on the possibility of observing the transition in electrolyte systems.

\section{Debye-H\"uckel Theory}\label{DHsection}

In this section we recap the full Debye-H\"uckel theory of the spin ice `magnetolyte' as given by Kaiser \emph{et al}.~\cite{Kaiser2018}.
The magnetolyte model of spin ice begins by positing that the relevant degrees of freedom in spin ice systems are captured by a dilute ensemble of $N$ singly- and $N_2$ doubly-charged magnetic monopoles with respective chemical potentials $\mu<0$ and $\mu_2=4\mu$. These may occupy any of the $N_0$ sites of the diamond lattice with nearest-neighbour distance $a$ and they interact with each other via a magnetic Coulomb potential, giving a Hamiltonian of the form
\begin{equation}\label{EQ: Hamiltonian}
    \mathcal{H} = \frac{1}{2} \sum_{i\neq j} \frac{\mu_0 q_i q_j}{4\pi r_{ij}}-\mu N - \mu_2 N_2
\end{equation}
where $q_{i,j} \in \{0, \pm Q, \pm 2Q \}$ with $Q$ a material-dependent elementary magnetic charge and $r_{ij}$ the separation between sites $i$ and $j$.

The behaviour of such an ensemble may be described by the free energy per site
\begin{equation}\label{Eq:FreeE}
    f = u_{\rm C} -\mu n - \mu_{2} n_{2} - sT
\end{equation}
where $u_{\rm C}$ is the Coulomb energy per site, $n$ and $n_2$ are the respective site densities of singly- and double-charged monopoles, $s$ denotes the entropy per site and $T$ denotes the temperature of the ensemble. The challenge is in determining the equilibrium value of $u_{\rm C}$, since the long-range Coulomb interaction acts between every pair of magnetic charges.

Debye-H\"uckel theory tackles this problem by considering how spatial correlations between monopoles in the system affect the first term of Eq. \ref{Eq:FreeE}. It is useful here to introduce the quantity $u(a) = \frac{\mu_0 Q^{2}}{4\pi a}$ as the Coulomb energy of a pair of singly charged monopoles separated by one lattice constant.

The total Coulomb energy of the system is determined by calculating the average energy associated with each monopole at a given monopole density $\rho_{\rm I} = (n + 4n_2)/\tilde{v}$, where $\tilde{v} = 8a^{3}/3\sqrt{3}$ is the volume per site. Noting that in the vicinity of a given monopole one is more likely to find monopoles of opposite charge than of like charge, the linearised Poisson-Boltzmann equation may be solved to give a screened Coulomb potential which differs from the Coulomb term of Eq. \ref{EQ: Hamiltonian} by a factor of $\exp{(-r/l_{\rm D})}$, where
\begin{equation}\label{Eqn: DebyeLength}
l_{\rm D} = \sqrt{\frac{k_{\rm B}T}{\mu_0 Q^{2} \rho_{\rm I}}}
\end{equation}
is the Debye length. The screening limits the average Coulomb energy of each monopole to that of the interaction between the monopole and its immediate atmosphere, giving~\footnote{Note that this corrects an error in the expression given for the Coulomb energy in Equation 13 of Ref.~\cite{Kaiser2018}, which is too small by a factor of four; the error does not propagate into the results of that paper.}
\begin{equation}\label{Eqn: UCoulomb}
u_{\rm C}^{\; \rm DH} = - \frac{2k_{\text{B}}T}{3\pi \sqrt{3}} \left[
\ln \left( 1+\frac{a}{l_{\text{D}}} \right) - \left( \frac{a}{l_{\text{D}}} \right) + \frac{1}{2} \left(\frac{a}{l_{\text{D}}}\right)^{2} \right]\\
\end{equation}

To complete the expression for the free energy, the entropy per site for spin ice may be expressed in a low-density approximation as 
\begin{equation}\label{entropy}
\begin{split}
        s = -k_{\text{B}} \bigg\{ n \ln \left( \frac{n}{2} \right) + n_{2} \ln (2n_{2})\\
     + (1-n-n_{2}) \ln (1-n-n_{2}) \\
     + (1-n-n_{2}) \ln \left( \frac{2}{3} \right) \bigg\}.
\end{split}
\end{equation}
This approximate entropy expression returns negative values (rather than zero) when the density approaches unity. However in practice, the contribution of $-sT$ to the free energy is always rather small when the negative entropy occurs so that its impact on observables is negligible. Since the expression in Eq. \ref{entropy} accurately describes experimental data, we do not attempt to correct the negative entropy it assigns to high-density configurations, but simply note where it occurs.

From these equations and for a given chemical potential,  one can calculate the equilibrium density that minimises the free energy as
\begin{equation}
    n = \frac{\frac{4}{3} \exp \left( \beta\tilde{\mu}\right)}{1+\frac{1}{3} [4 \exp\left(\beta \tilde{\mu}\right)+\exp\left(\beta \tilde{\mu}_{2}\right)]}\\
\end{equation}
\begin{equation}
    n_2 = \frac{\frac{1}{3} \exp \left( \beta\tilde{\mu}_{2}\right)}{1+\frac{1}{3} [4 \exp\left(\beta \tilde{\mu}\right)+\exp\left(\beta \tilde{\mu}_{2}\right)]}
\end{equation}
where $\beta=1/k_{\rm B}T$ and $\tilde{\mu}, \tilde{\mu}_{2}$ are effective chemical potentials which depend on the Debye length and hence the monopole density as

\begin{equation}\label{DeltaDH}
\begin{gathered}
\tilde{\mu} = \mu+\Delta^{\rm DH},\;\; \tilde{\mu}_{2}=\mu_2+4\Delta^{\rm DH}\\
\Delta^{\rm DH} = k_{\rm B} T \frac{l_{\rm T}}{l_{\rm D}+a}.
\end{gathered}
\end{equation}
From here it is possible to find solutions iteratively, alternately adjusting the effective chemical potential and density until a self-consistent solution is obtained. 

The iteration will generally converge to a single free energy minimum, so if there is a double minimum in the free energy (as we expect for the first order transition), it is important to start the iteration at different initial densities, so that both minima can be found. The procedure that we settled on involved starting at a low temperature with initial densities set to zero, so that for the first iteration $\tilde{\mu}=\mu$ and $\tilde{\mu_2} = \mu_2$. This typically allows the iterative solution to converge within 10 steps. The temperature was then increased in steps of 10 mK, at each step using the converged $\tilde{\mu}_i$ and $n_i$ from the previous temperature as the starting point. Once a high temperature was reached, the system was cooled, again in small temperature steps, each time using the parameters from the previous step to start the iteration. In this way the cooling curve would give the absolute free energy minimum while the heating curve would follow a metastable minimum if there was one. Some tests showed that this procedure indeed located the correct (global) minima.

\section{Results}

\subsection{Monopole density and charge density} \label{density}

Using the physical quantities for dysprosium titanate ($a=4.34$ \AA\; and magnetic charge $Q=4.28 \times 10^{-13}$ Am) and a variable chemical potential, effective chemical potentials $\tilde\mu_i$ and monopole densities $n_i$ ($i = 1,2$ for singly- and doubly-charged monopoles respectively) were iterated to convergence as described above.

\begin{figure}[ht]
    \centering
    \includegraphics[height=6.7cm]{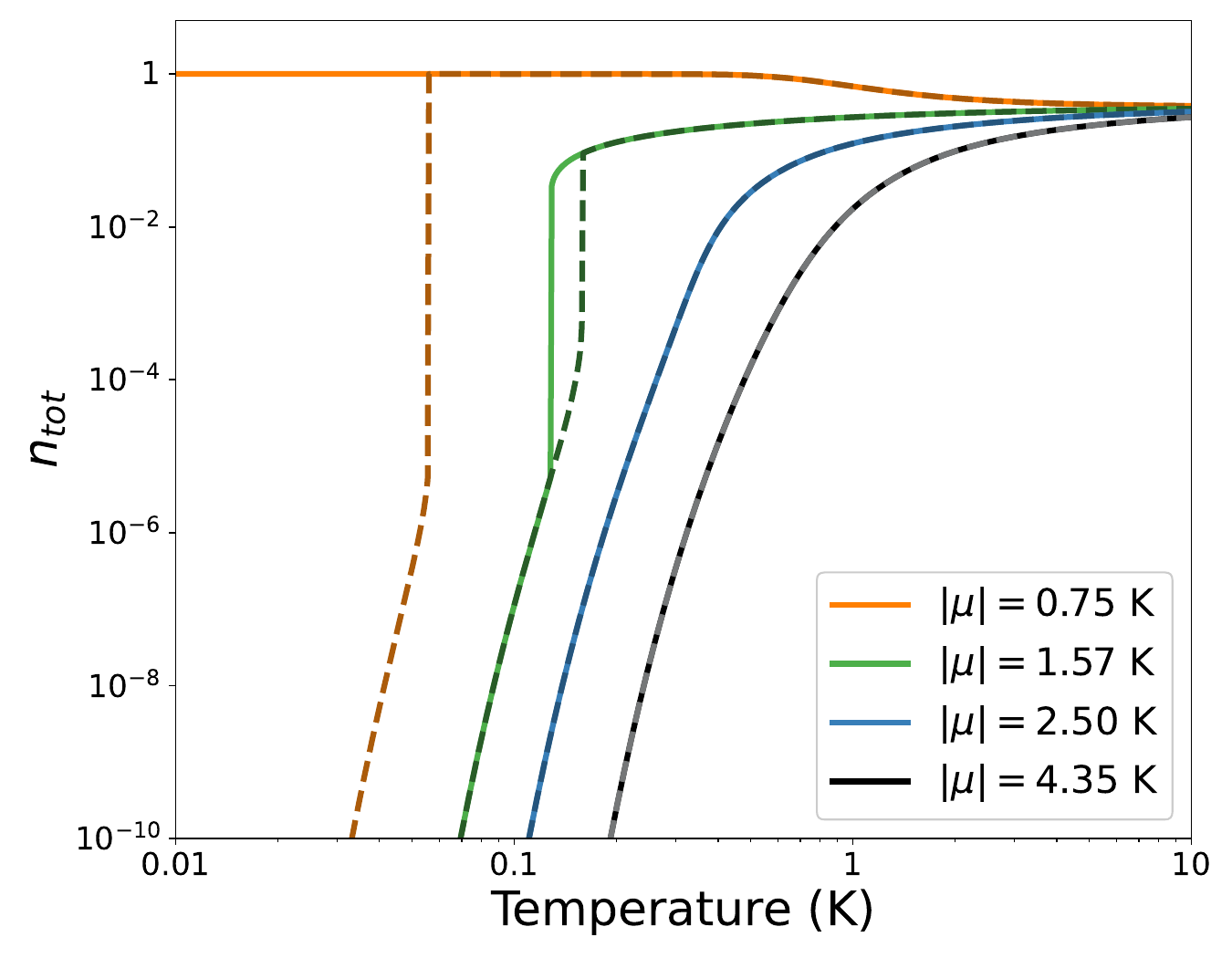}
    \caption{
 Total monopole density $n_{\rm tot} = n +n_2$ as a function of temperature for varying chemical potentials $\mu$ with $\mu_2=4\mu$. Solid lines represent cooling curves (representing the true equilibrium state) and dashed lines represent heating curves (metastable where they deviate from the full lines). There is a first order transition and associated hysteresis when $1.6 \;{\rm K} \gtrsim |\mu| \gtrsim 1.3$ K.}
    \label{fig:transition}
\end{figure}

Fig. \ref{fig:transition} illustrates the resulting curves of total monopole density per site, $n_{\rm tot} = n_1+n_2$,  versus temperature. First order phase transitions are observed at bare chemical potentials of $|\mu| \lesssim 1.6$ K, with larger magnitude transitions at smaller temperatures. The heating/cooling cycle reveals hysteresis in these transitions, with the cooling curve generally finding the stable free energy minimum (full lines in Fig.1). Our results, however, indicate three regimes, depending on chemical potential: (i) For small $|\mu| \lesssim 1.3$ K, the lowest temperature state has equilibrium single-monopole densities approaching unity, while the free energy minimum found by heating is metastable up to the transition. The unit density state is reached at temperatures much higher than the first order transition, which only occurs for the metastable heating curve: that is, there is no equilibrium first order transition, but the first order transition does persist as a metastable feature. (ii) For $1.6 \;{\rm K} \gtrsim |\mu| \gtrsim 1.3$ K, the single monopole equilibrium density falls to zero at low temperature and the heating curve is metastable only in the vicinity of the transition. Here there is an equilibrium first order transition between high and low density monopole states with thermal hysteresis in the total monopole density. (iii) At $|\mu| \gtrsim 1.6$ K there is a unique free energy minimum which is followed in both the heating and cooling curves, and hence no first order transition. 

\begin{figure}[ht]
    \centering
    \includegraphics[height=6.5cm]{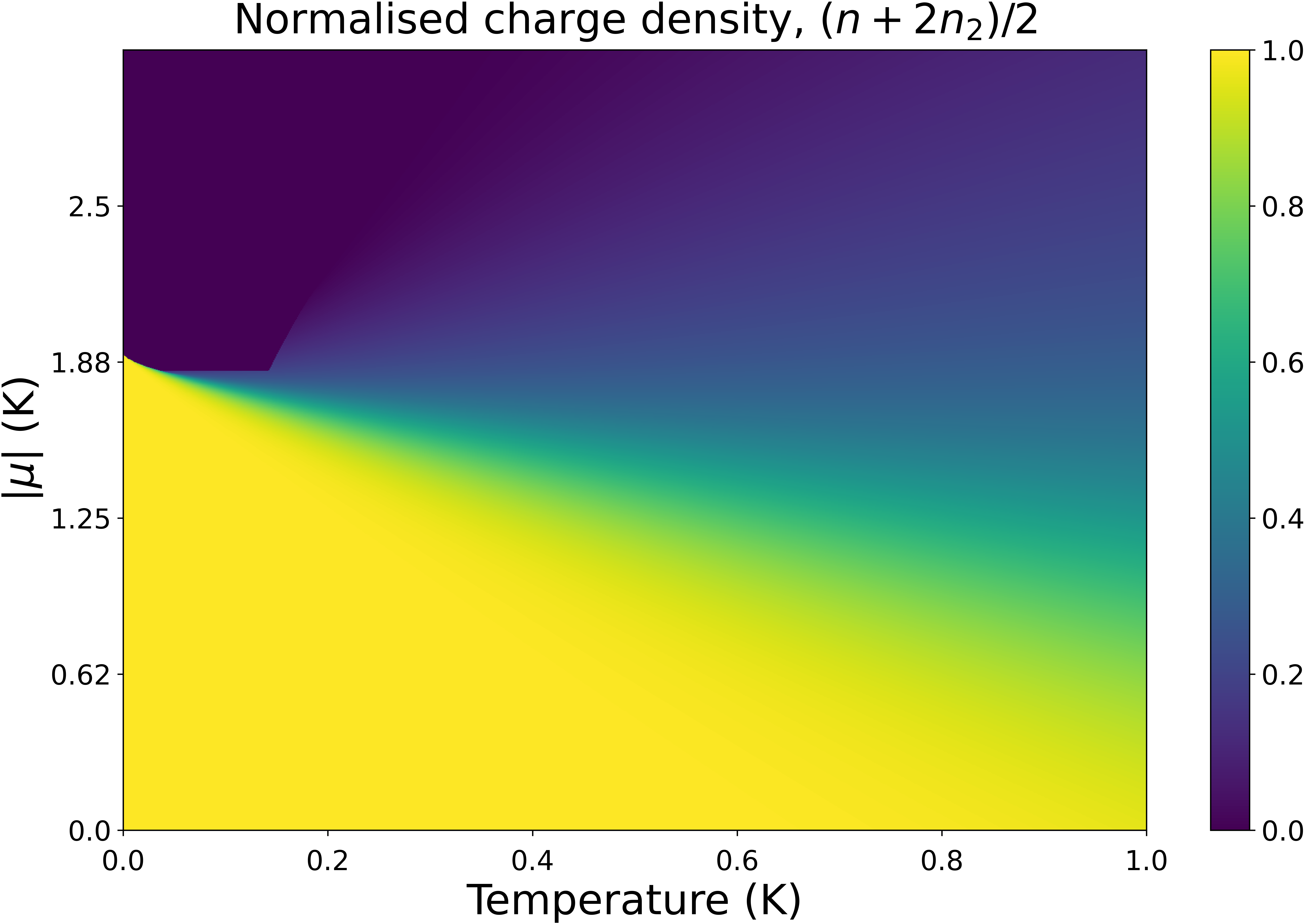}
    \caption{The normalised equilibrium charge density $(n+2n_2)/2$, computed as a function of $|\mu|$ and $T$, with standard material parameters for Dy$_2$Ti$_2$O$_7$ and $\mu_2=4\mu$. Note the line of first order phase transitions near to $|\mu| = 1.5$ K and $T \rightarrow 0$.}
    \label{fig:equilibriumPhaseDiagram}
\end{figure}

The evolution of the normalised charge density per site $(n+2 n_2)/2$  with temperature and chemical potential directly reflects the properties described above. Fig. \ref{fig:equilibriumPhaseDiagram} shows the equilibrium charge density as a function of $|\mu|$ and $T$, with a short line of first order phase transitions near to $|\mu| = 1.5$ K and $T \rightarrow 0$.

\subsection{Specific heat}

\begin{figure}[ht]
    \centering
    \includegraphics[height=6.7cm]{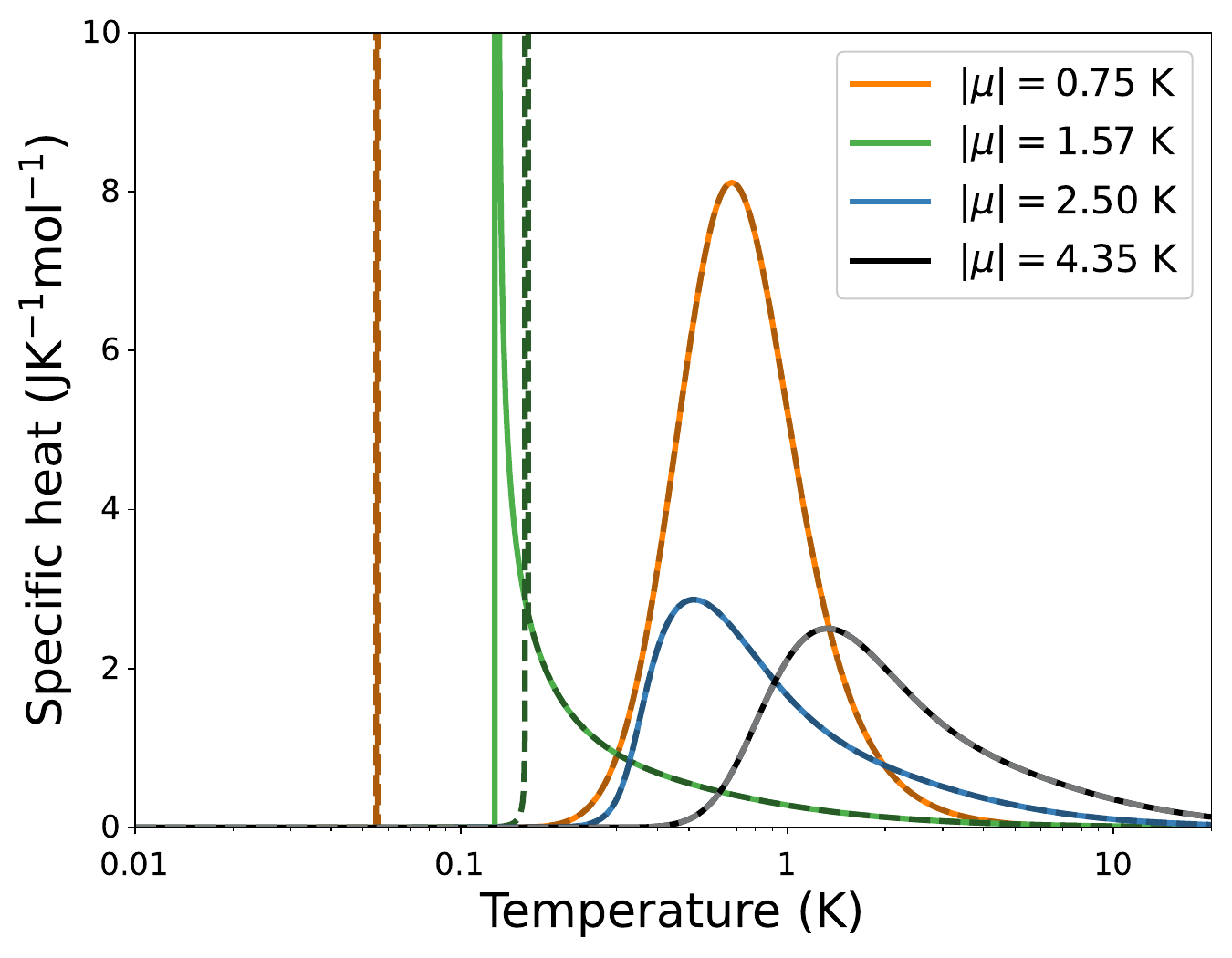}
    \caption{Specific heat as a function of temperature for varying monopole chemical potential $\mu$, showing three distinct regimes as in Fig. \ref{fig:transition}. The cooling (equilibrium) curves are solid, with the heating (metastable) curves dashed.}
    \label{fig:specific heat}
\end{figure}

Computing the specific heat as a function of temperature (Fig. \ref{fig:specific heat}) sheds further light on the nature of the three regimes (i) - (iii) identified in the previous section. In regime (i), the cooling (equilibrium) curve features a single broad peak associated with the crossover from a single-monopole dominated limit at high temperatures to a double-monopole dominated limit at low temperatures. The large area under this peak is a result of the negative entropy assigned to configurations with monopole densities approaching unity, discussed where Eq. \ref{entropy} was introduced. At lower temperatures, the heating (metastable) curve has a second very sharp peak where the double-monopole density discontinuously jumps by many orders of magnitude. The ``monopole density inversion'' where the $n_2$ becomes greater than $n_1$ is discussed further in the following subsection.

In regime (ii) the specific heat diverges when both the single- and double-monopole densities abruptly increase with the temperature by several orders of magnitude, with the thermal hysteresis in the monopole densities reflected in a shift in the temperature at which the divergence occurs. Close to these critical temperatures the specific heat has the asymmetric form characteristic of a mean field transition, as might be expected in this effective field model. In contrast to regime (i), $n_1$ remains greater than $n_2$ for all temperatures.

In regime (iii) with $|\mu| > 1.6$ K the heating and cooling curves are identical, with broad, continuous peaks in specific heat with larger $|\mu|$ shifting the peaks to higher temperatures. These peaks are the familiar Schottky anomalies associated with single monopole activation as the temperature approaches their chemical potential. As for regime (ii), $n_1$ remains greater than $n_2$ for all temperatures.

\subsection{Monopole density inversion}

\begin{figure}[ht]
    \centering
    \includegraphics[height=6.7cm]{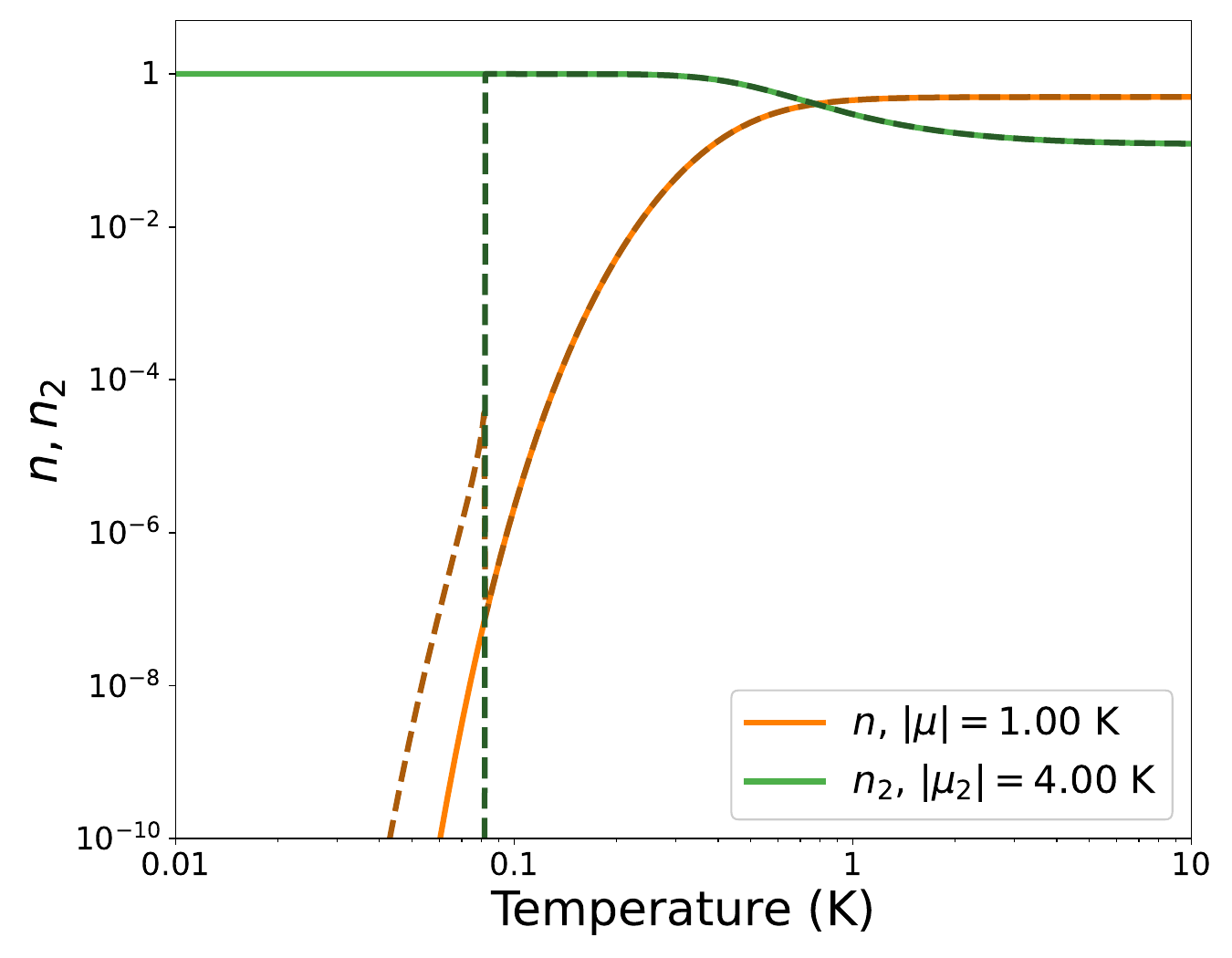}
    \caption{Single ($n$) and double ($n_2$) monopole site densities as a function of temperature for $|\mu| = 1.00$ K with $\mu_2=4\mu$. Solid lines represent cooling curves at equilibrium while dashed lines indicate heating curves which are metastable where they do not coincide with the cooling curves.}
    \label{fig:inversion}
\end{figure}

In the canonical model of classical spin ice, the chemical potential for doubly charged monopoles is always four times that for singly charged monopoles. Hence in the low density and low temperature limit, for larger chemical potentials, $n_2 \ll n_1$, while in the high temperature limit, $n_2 = 1/8$ and $n_1 = 1/2$. However, we discovered that below the first order transition at $|\mu|\approx 1.5$ K, doubly-charged monopoles dominate and displace singly charged monopoles, then remaining dominant while the system remains in a high monopole density state. This behaviour is illustrated in Fig. \ref{fig:inversion} for $|\mu_1| = 1$ K, i.e. in regime (iii) above.
At equilibrium (full line in figure) double monopoles smoothly start to dominate below $T \sim 1$ K.  Although there is no density jump at equilibrium, there is a metastable first order transition in the heating curve at $T \sim 0.1$ K, where the relative site densities of single and double charge monopoles invert to reach the equilibrium values.

For a more general case (not shown), it is interesting to relax the constraint that $\mu_2 = 4 \mu$. For $\mu_2 < 4 \mu$ this allows the double monopoles to dominate for a larger temperature range immediately above the first-order transition, while for $\mu_2 \gg 4\mu$ the inversion is suppressed and single monopoles remain dominant in both the low- and high-density states. We discuss in Section \ref{StaggeredPot} how a staggered interaction relevant to spin ice iridates can lead to ``dressed'' chemical potentials that effectively break the constraint $\mu_2 = 4 \mu$.

\subsection{Transitions in electrolytes} \label{electro}

Since the Debye-H\"uckel picture outlined above can in principle be applied to electrolytes in general, it is natural to ask whether the first-order transition described in this work is relevant to other systems.

\begin{figure}[ht]
    \centering
    \includegraphics[height=6.8cm]{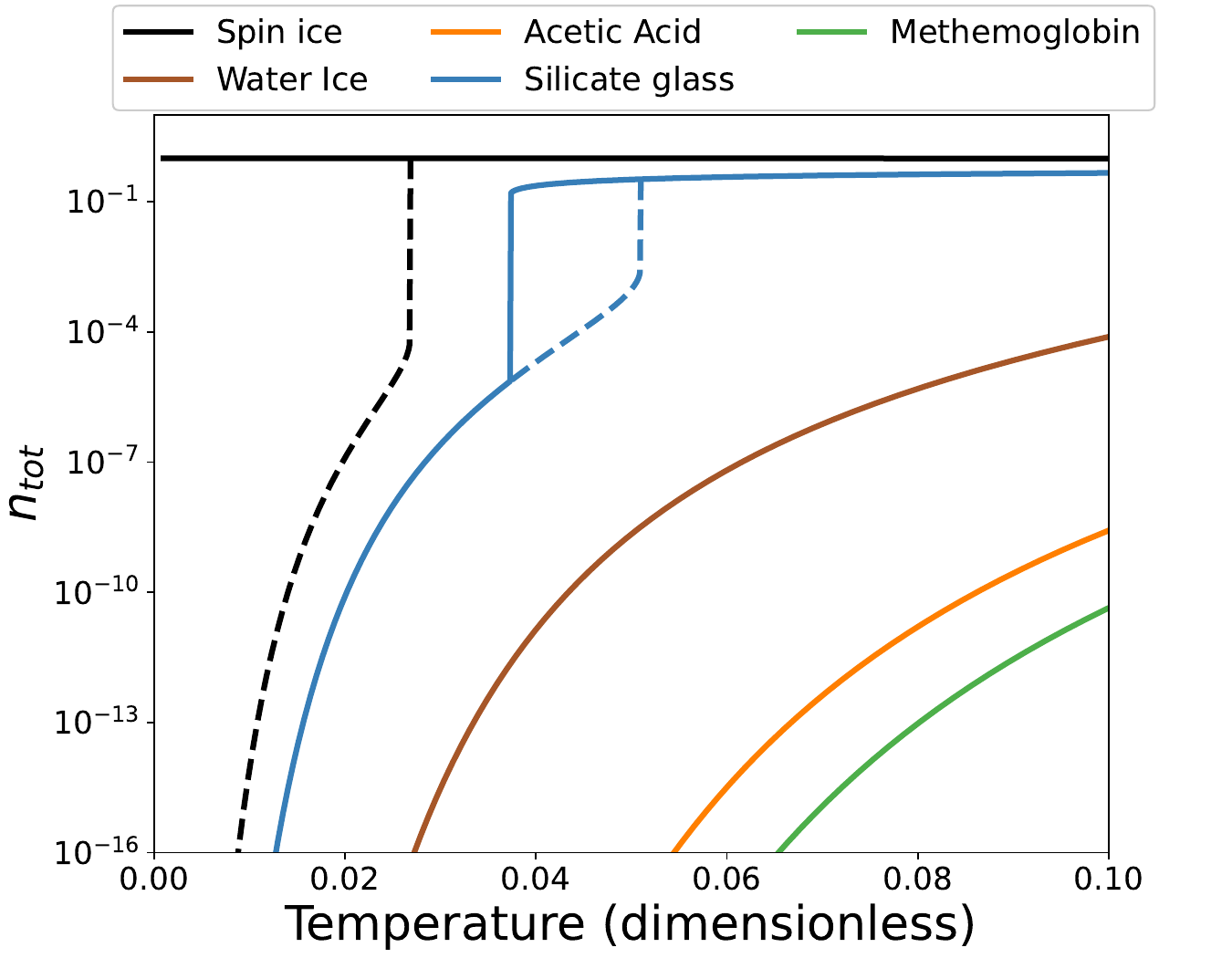}
    \caption{Comparison of monopole densities against temperature (in the reduced units of Ref. \cite{Kaiser2013} for several ice-type and electrolyte systems. Note that `spin ice' here is a hypothetical system with  $|\mu| = 1.00$ K and otherwise, ${\rm Dy_2Ti_2O_7}$ parameters, while other systems have realistic parameters as estimated in Ref. \cite{Kaiser2013}. With the exception of the silicate glass~\cite{Ingram1980}, it seems that the equilibrium phase transition is not generally available in these systems. Here the dimensionless temperature is defined as $T^{\ast} = k_{\rm B} T/u(a)$ where $u(a)$ is defined in Section \ref{DHsection}.  }
    \label{fig:electrolytes}
\end{figure}

One difference between spin ice and an  ordinary electrolyte is that spin ice has a structured vacuum for charge excitations, which is reflected in the details of the entropy, Eqn. \ref{entropy}. Specialising to the case of single charges (density $n$),  a symmetric lattice electrolyte may be described by the equations used here, provided Eqn. \ref{entropy} is replaced by the primitive electrolyte entropy~\cite{Kaiser2018}:
\begin{equation}
S_{\rm e} =-k_{\rm B} N_{\rm 0}\left[ n \log (n/2)+(1-n)\log(1-n)\right].
\end{equation}
It was confirmed that the first order transition is maintained when this entropy expression is used in place of Eqn. \ref{entropy}. To give some sense of the magnitudes involved we refer to the table of electrolyte parameters given by Kaiser \emph{et al.} in Ref. \cite{Kaiser2013}. Results for various electrolytes were generated using these parameters and the primitive entropy, including the case of water ice (where the ice entropy was used) and spin ice with  as a comparison. These results are illustrated in Fig. \ref{fig:electrolytes}. There is a first order transition for silicate glass ($\rm Na-Ca-SiO_2$) and a metastable one for methemoglobin. Whether or not the real systems would display such transitions is discussed subsequently.

\subsection{Breakdown of Debye-H\"uckel theory} \label{breakdown}

\begin{figure}[ht]
    \centering
    \includegraphics[width=0.45\textwidth]{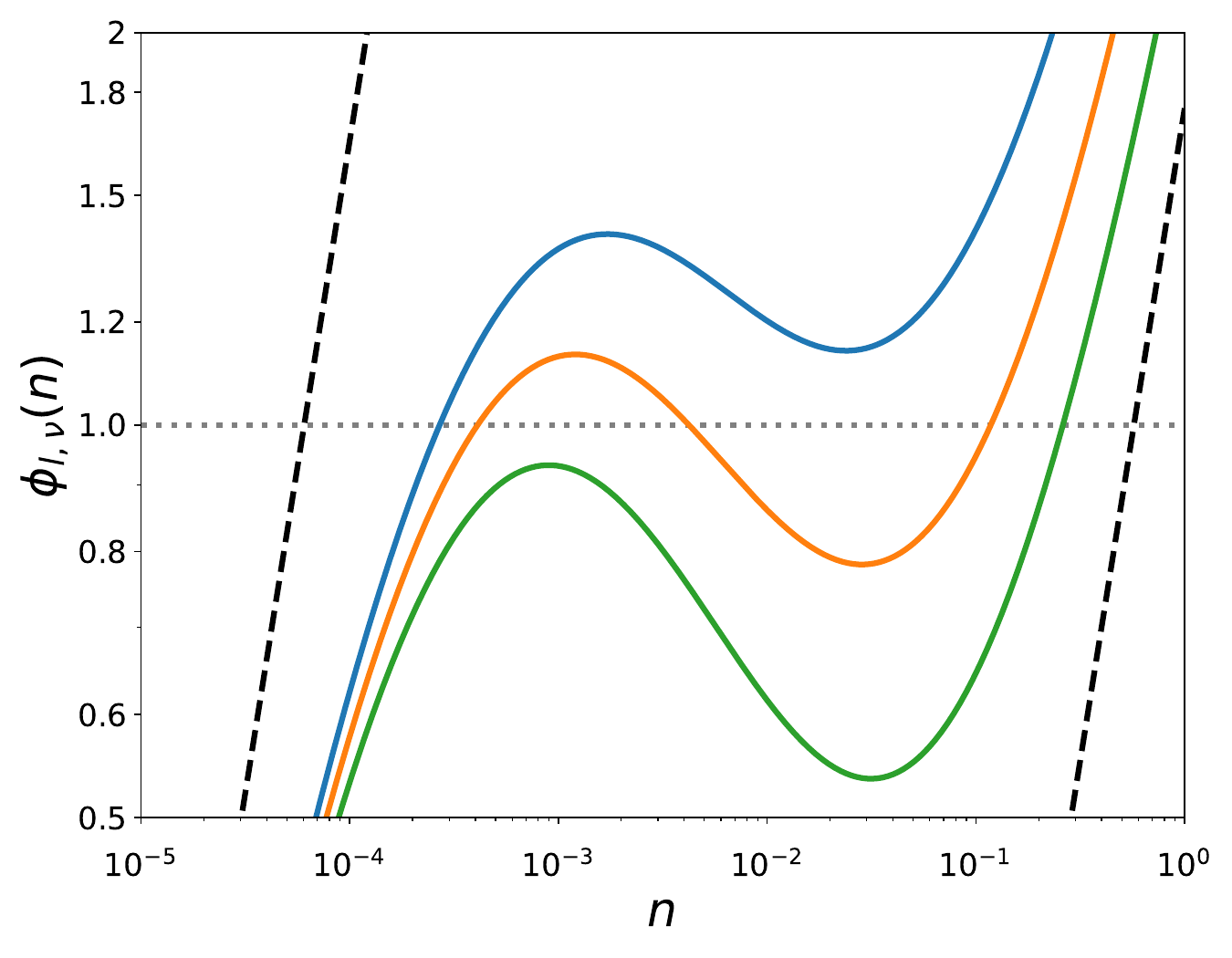}
     \includegraphics[width=0.45\textwidth]{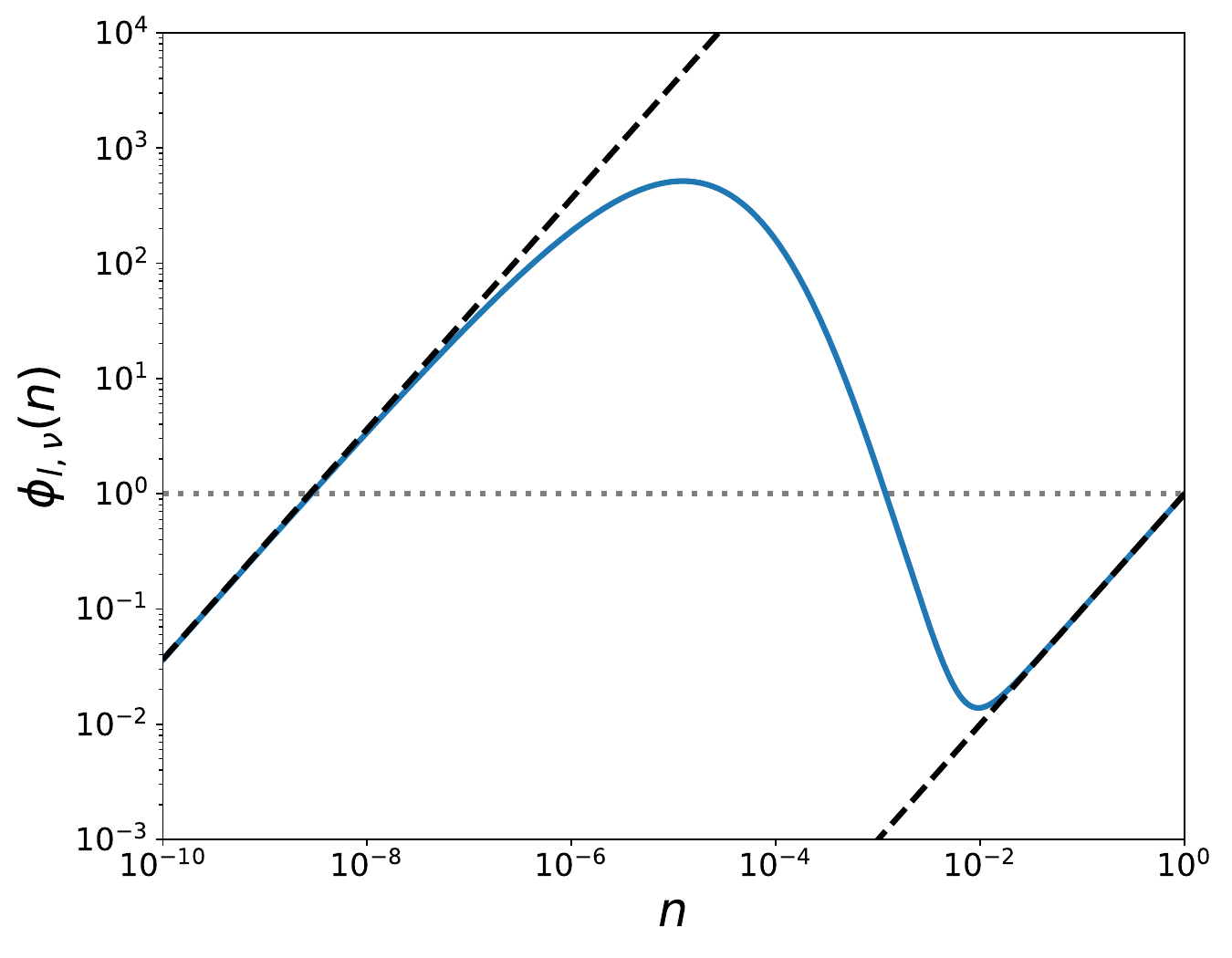}   
    \caption{The Debye-H\"{u}ckel free energy has equilibria where the function $\phi_{l, \nu}(n) =  1$ (Eqn. \ref{sol})). Here we plot $\phi_{l, \nu}(n)$ versus density $n$ for different values of $l$ and $\nu$, while the straight dashed lines represent the limiting forms for small and large $nl$ (left and right, respectively) (Eqn. \ref{sol2}).  Equilibrium solutions for $n$ occur where the solid curves intersect the dotted line at $\phi = 1$ (see Eqn. \ref{sol}). The upper plot has $\nu = 10$ and (curves, left to right) $l = 9,9.5,10$; the lower plot has $\nu = 20,\; l = 30$.}
    \label{fig:phi}
\end{figure}

The question arises, do the transitions observed occur in a parameter range where Debye-H\"uckel theory gives a valid description of the Coulomb fluid? To explore this question, we specialise to the single-charge case and define parameters $l = l_{\rm T}/a$,  $\nu = -\mu/k_{\rm B}T$. Note that $l$ may be expressed equivalently as $l = u(a)/2k_{\rm B}T$, where $u(a) = \mu_0Q^2/4\pi a$ as before. It is then easily shown that there are turning points in the free energy when
\begin{equation}\label{sol}
\phi_{\nu,l}(n) \equiv n \left(1+ (3/4) e^{\nu-\frac{l \sqrt{n l}}{\sqrt{n l}+c}}\right)  = 1
\end{equation}
where  $c = 1/(\sqrt{\pi}\, 3^{3/4}) \approx 0.2475$. 
We can study this equation for the case of $nl \ll c $ and $nl \gg c$ respectively:
\begin{equation}\label{sol2}
n \left(1+ (3/4) e^{\nu} \right) = 1~~~~~~~~~n \left(1+ (3/4) e^{\nu-l}\right)  = 1.
\end{equation}
Each of these equations has only one solution. However, three solutions, that we interpret as two minima and a maximum in the free energy, can arise when there is a crossover between the two limiting forms (Fig. \ref{fig:phi}). 
From inspection, this requires $l\sim \nu$ which evaluates to $(u(a)/2) \approx |\mu|$, that is, half the Coulomb energy ($u\approx 3$ K) of a pair at contact -- as we have observed, there is indeed an equilibrium transition for $|\mu| \approx 1.5$ K.  In addition, it is clear from inspection that $l, \nu$ need to be quite large, of order 10 or more, for there to be multiple minima.

On the other hand, the formal criterion for Debye-H\"uckel theory to be valid is that the total magnetostatic energy in a potential $\Phi$ dominates the thermal energy scale, $1 \gg Q\Phi/ k_{\rm B} T =  2 \Delta_{\rm DH} (l_{\rm D}/a)$, which becomes : 
\begin{equation}
1\gg \frac{2 \sqrt{\pi } l}{3^{3/4} \sqrt{\pi n l}+1}. 
\end{equation}
The right hand side of this equation is a monotone decreasing function of $n$ so we can identify $l^{\rm max} \approx 1.82$ as the solution for $n=1$. There is only one solution to Eqn. \ref{sol} for all values of $\nu$ for $l \le l^{\rm max}$, suggesting that the first order transition occurs in a parameter regime where Debye-H\"uckel theory breaks down.

One can expect the Bjerrum correction for bound pairs, as applied in Ref.\cite{Kaiser2013}, to extend the range of validity of the theory to values of $l$ rather larger than $l_{\rm max}$, but as illustrated in Fig. \ref{fig:phi}, the length $l$ is always large at the transition and it seems unlikely that Debye-H\"uckel-Bjerrum theory can ever be strictly valid in this regime. Physically, such large values of $l$ indicate a tendency to pairing, or strong correlation, driven by the Coulomb interaction. But more realistically, while the Debye-H\"uckel theory can only describe a fluid to fluid transition, for $|\mu| \le \mu_{\rm c}$ (i.e. the regime of interest) it has been shown that the system is unstable against crystallisation, driven by a favourable Madelung energy~\cite{Raban2019}, and it may be reasonable to interpret the transition in the linearised Debye-H\"uckel theory as a remnant of this.

\section{Effect of a Staggered Potential}\label{StaggeredPot}

In light of an apparent relation to monopole crystallisation transitions at low temperatures, it is interesting to explore whether Debye-H\"{u}ckel theory can be brought into contact with other models of spin ice that predict crystallisation. These models may be separated into monopole-conserving models \cite{Borzi2013,Guruciaga2014} in which charge-ordering is observed at high densities and models associated with the literature of spin ice fragmentation \cite{BrooksBartlett2014,Petit2016,Raban2019,Lhotel2020, raban2022violation}. The most general case of the latter, as described by Raban \emph{et al.}~\cite{Raban2019}, supplements the Hamiltonian of the dipolar spin ice model in the dumb-bell picture with a staggered interaction, giving
\begin{equation}
	\mathcal{H} = \frac{u(a)}{2} \sum_{i \neq j} \frac{a}{r_{ij}}\hat{n}_{i}\hat{n}_{j} -\mu \sum_{i} \hat{n}_{i}^{\; 2} - \Delta \sum_{i} (-1)^{i} \hat{n}_{i}
\end{equation}
where $u(a) = \frac{\mu_0 Q^{2}}{4 \pi a}$ (see above), $\hat{n}_{i}=\pm 1, \pm 2$ is the monopole occupation number for the $i$th site and $\Delta>0$ is the strength of a staggered field. The first two terms are the same Coulomb interaction and chemical potential we have treated already, while the staggered field introduced here promotes ordered monopole states by shifting the energy cost of introducing monopoles from $\mu \rightarrow \mu \pm \Delta$ for isolated single monopoles and from $4\mu \rightarrow 4\mu \pm 2\Delta$ for isolated double monopoles, with the sign depending on whether the monopole is on an odd or even site. Such staggered potentials have been explored in the context of Ho$_{2}$Ir$_{2}$O$_{7}$ (HIO) \cite{Lefrancois2017}, in which both Ho$^{3+}$ and Ir$^{4+}$ ions have a net magnetic moment. The Ir$^{4+}$ ions, which sit on a pyrochlore lattice that intersects with the Ho$^{3+}$ pyrochlore structure, undergo an ordering transition to an antiferromagnetic all-in-all-out arrangement at temperatures well above the Coulomb scale. The internal magnetic fields generated by the  Ir$^{4+}$ sublattice thus have a staggered structure that promotes staggered order in the Ho$^{3+}$ moments.

\begin{figure}[ht]
    \centering
    \includegraphics[width=0.45\textwidth]{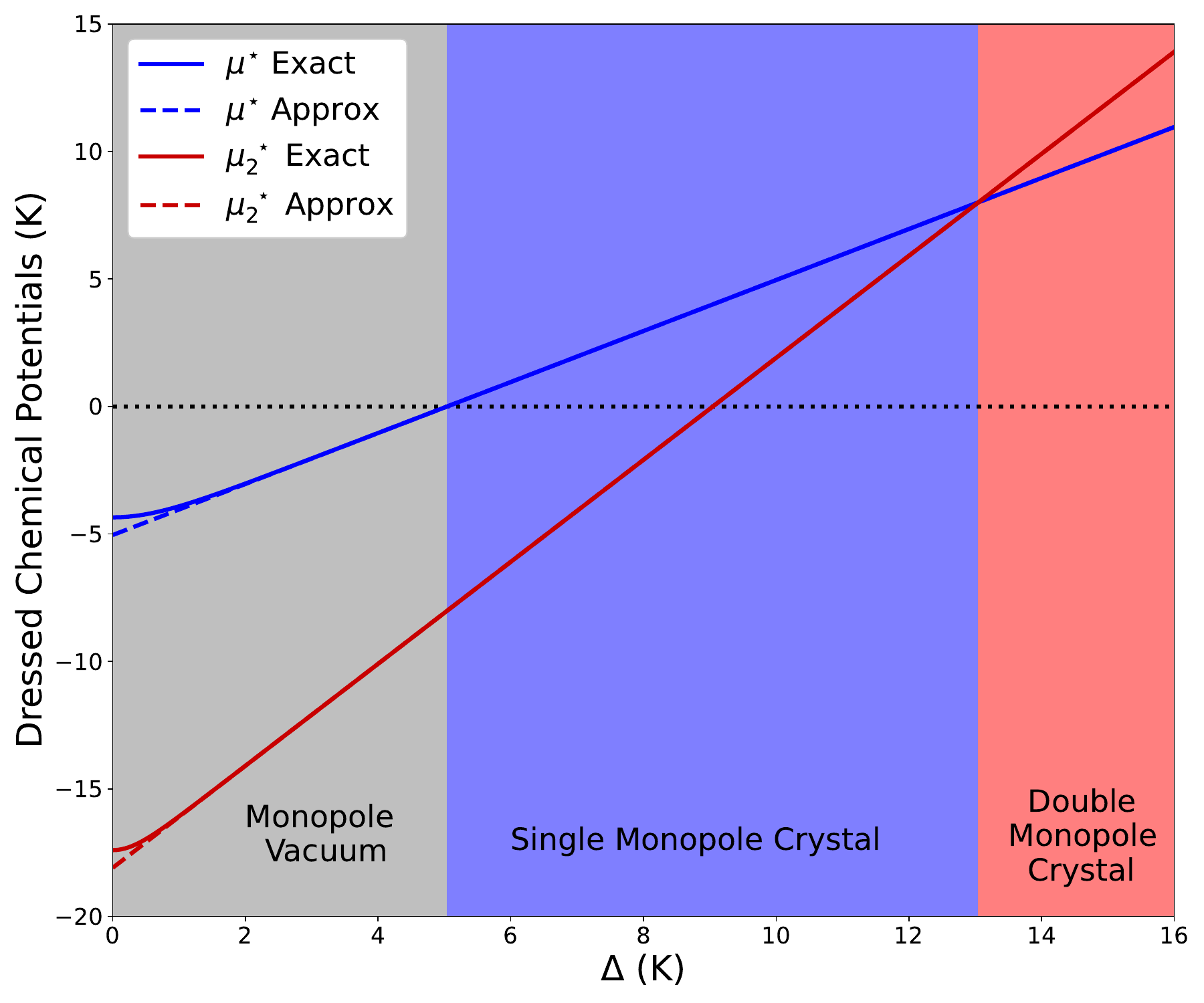}
    \caption{The exact (solid) and approximate (dashed) forms of the dressed potentials $\mu^{\star},\; \mu_{2}^{\;\star}$ (Eqs. \ref{Eq:MuStar} and \ref{Eq:MuStar2}, respectively) as a function of the staggered potential $\Delta$ with $\mu=-4.35$ K and $\beta=1$ K$^{-1}$. The points where $\mu^{\star}$ crosses zero and where $\mu_{2}^{\;\star}$ becomes greater than $\mu^{\star}$ form the zero-temperature phase boundaries between the monopole vacuum and single- and double-monopole crystal phases for the non-interacting limit ($Q\to0$).}
    \label{fig:DressedMuZeroQ}
\end{figure}

As well as distinguishing alternate lattice sites, the staggered field term will also tend to increase the monopole population at a given temperature since $e^{\beta(\mu + \Delta)} + e^{\beta(\mu - \Delta)} \geq 2 e^{\beta\mu}$ for all $\Delta \geq 0$. We can incorporate this latter effect into our Debye-H\"{u}ckel picture by introducing a ``dressed'' chemical potential $\mu^{\star}$ that averages the contributions from odd and even sites so that 
\begin{equation}
	e^{\beta \mu^{\star}} = \frac{1}{2} \left( e^{\beta(\mu + \Delta)} + e^{\beta(\mu - \Delta)} \right).
\end{equation}
This may be solved to give an expression for the (temperature-dependent) dressed chemical potential in terms of bare chemical potential and the staggered field strength
\begin{equation}\label{Eq:MuStar}
	\mu^{\star} = \mu + \frac{\ln \left( \cosh (\beta \Delta) \right)}{\beta}
	\approx \mu + \Delta - \frac{\ln(2)}{\beta}
\end{equation}
where the approximate form is obtained by expanding the $\cosh (\beta \Delta)$ term for large $\beta\Delta$ (in practice, the approximate form converges rapidly and is already highly accurate for $\beta\Delta\gtrsim 1$ as seen in Fig. \ref{fig:DressedMuZeroQ}). We see immediately that $\mu^{\star}>\mu \;\; \forall \beta,\Delta$ so that monopole populations will be greater than without the staggered potential, as expected. We similarly obtain
\begin{equation}\label{Eq:MuStar2}
		\mu_{2}^{\;\star} = 4\mu + \frac{\ln \left( \cosh (2\beta \Delta) \right)}{\beta} \approx 4\mu + 2\Delta - \frac{\ln(2)}{\beta}
\end{equation}
so that varying the strength of the staggered field has the effect of allowing the ratio $\mu_{2}^{\;\star} / \mu^{\star}$ to take arbitrary values.

In the non-interacting limit ($Q\to0$), we may readily use these approximate forms to determine the low-temperature phase boundaries between the vacuum, single-monopole crystal and double-monopole crystal phases (Fig. \ref{fig:DressedMuZeroQ}). The low-density phase occurs when $\mu_{2}^{\;\star} < \mu^{\star} < 0$, so that it is difficult to excite any monopoles. As $\Delta$ is increased, $\mu^{\star}$ becomes positive when $\Delta > -\mu+\ln(2)/\beta$ which defines the first phase boundary. The double-monopole crystal phase is reached when $0 < \mu^{\star} < \mu_{2}^{\;\star}$, corresponding to the phase boundary $\Delta=-3\mu$. At zero temperature, these phase boundaries are in exact agreement with those identified by Raban \emph{et al.} \cite{Raban2019,RabanThesis} based on an analogy with the $S=2$ Blume-Capel model.

\begin{figure}[ht]
    \centering
    \includegraphics[width=0.5\textwidth]{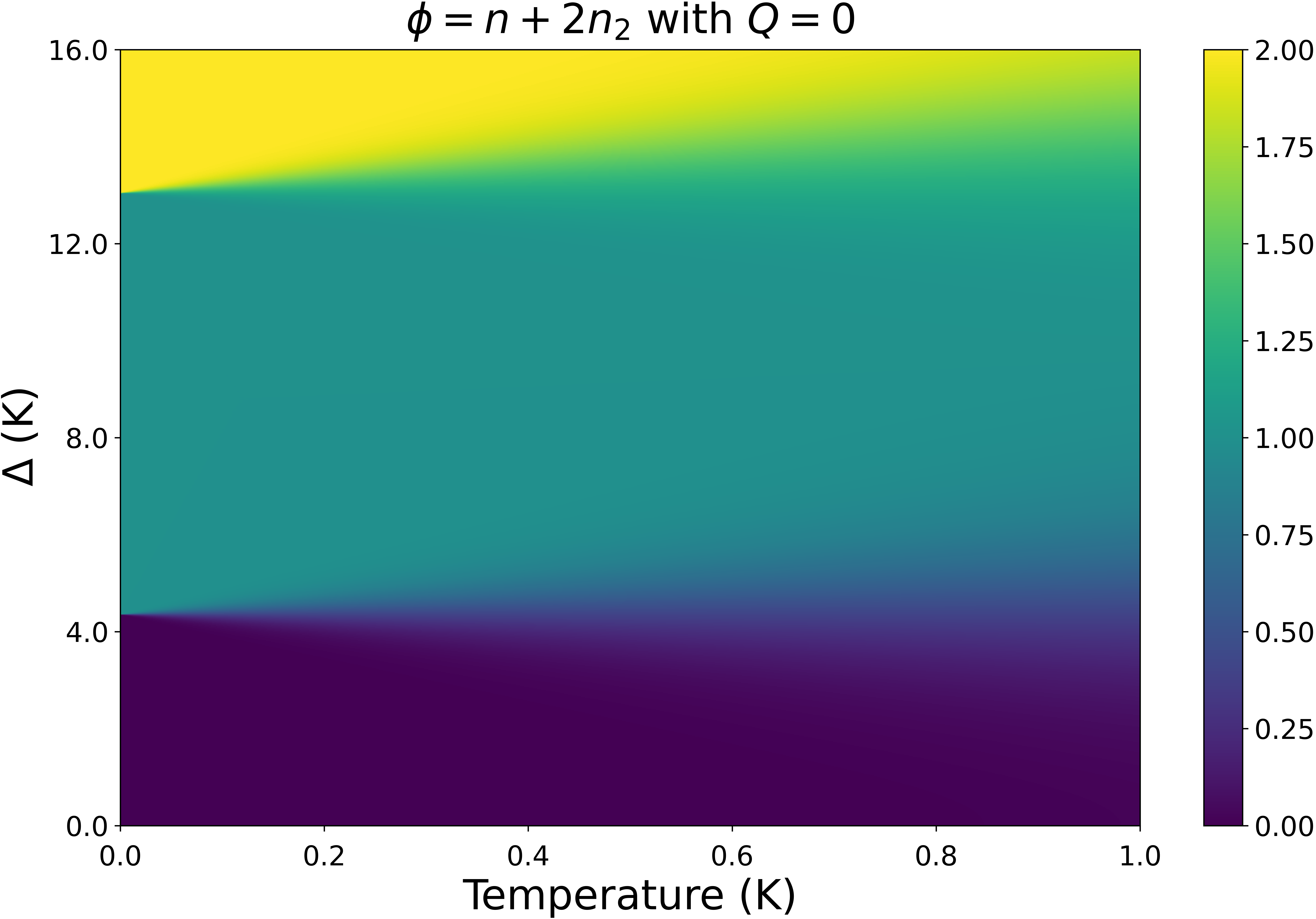}
     \includegraphics[width=0.5\textwidth]{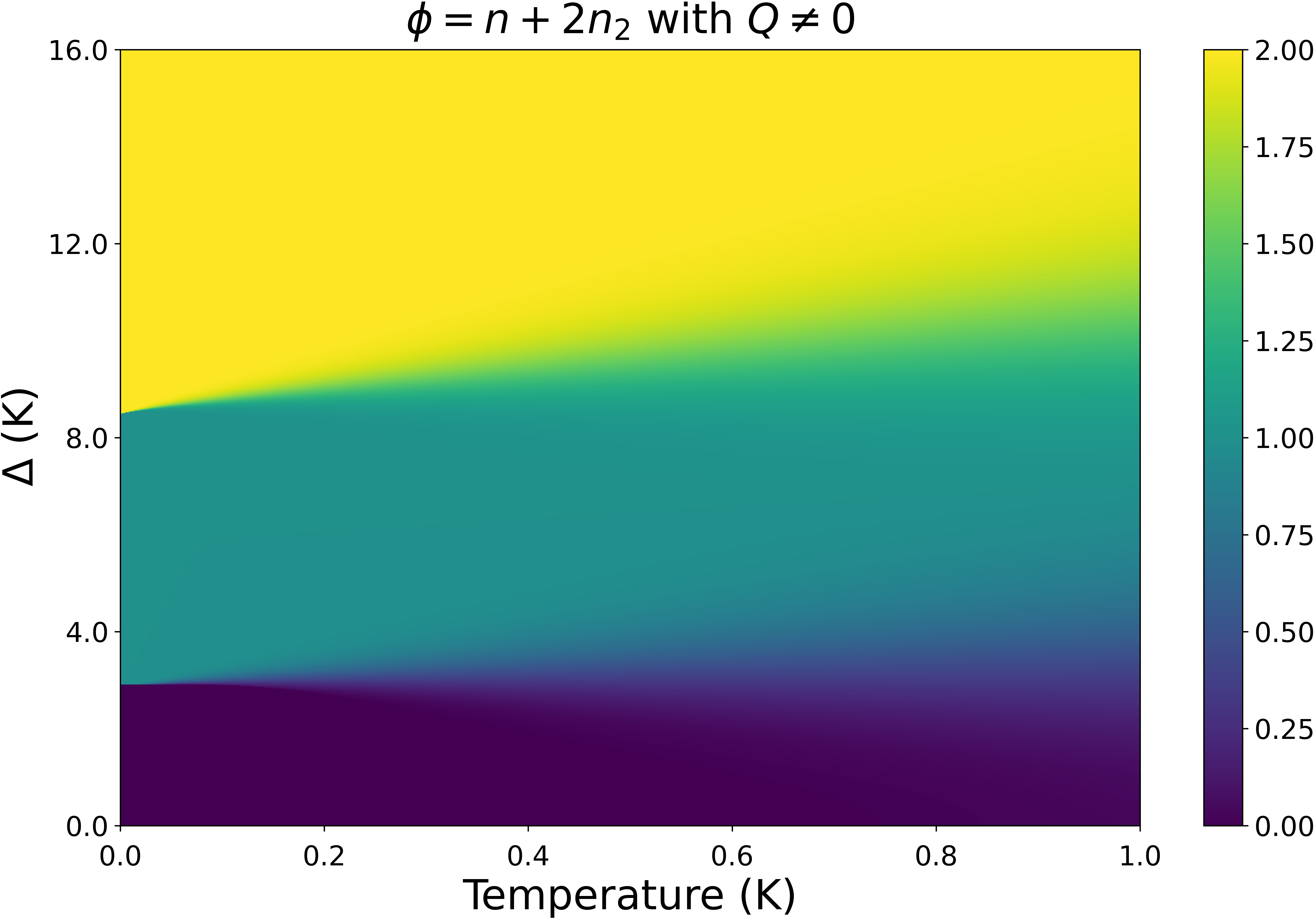}   
    \caption{Monopole charge density $\phi=n+2 n_2$ in the non-interacting limit (above) and in the case of interacting monopoles (below) calculated using material parameters for DTO with $\mu=-4.35$ K and an additional staggered field $\Delta$. Low-temperature phase boundaries in the non-interacting limit agree with the crossing points identified in Fig. \ref{fig:DressedMuZeroQ}. In the interacting case, Coulomb interactions suppress the single-monopole crystal phase (where $\phi\approx 1$) and shift the low-temperature phase boundaries down to lower values of $\Delta$ given by Eqns. \ref{Eq:ZeroTempVac1MP} and \ref{Eq:ZeroTemp1MP2MP}, in remarkable agreement with previous results by Raban \emph{et al.}~\cite{Raban2019}.}
    \label{fig:DeltaPhaseDiagrams}
\end{figure}

For the interacting case ($Q\neq0$), charge screening reduces the magnitude of the effective chemical potentials as seen in previous sections, shifting the phase boundaries down to lower $\Delta$. Figure \ref{fig:DeltaPhaseDiagrams} shows the equilibrium charge densities as a function of temperature and the staggered potential $\Delta$ for both the non-interacting and interacting case. To determine the effect of incorporating charge screening on the phase boundaries  we may use the zero-temperature correction to the dressed chemical potential, $\Delta^{\rm DH}(T=0)=\frac{\mu_0 Q^{2}}{8 \pi a}$, which is independent of the monopole  densities. We may then obtain the zero-temperature phase boundaries as before by looking at the hierarchy of $0$, $\tilde{\mu}^{\star}$ and $\tilde{\mu}_{2}^{\;\star}$, giving
\begin{equation}\label{Eq:ZeroTempVac1MP}
    \Delta_{\rm Vac \leftrightarrow 1MP} = - \left( \mu + \frac{\mu_0 Q^{2}}{8 \pi a} \right),
\end{equation}
\begin{equation}\label{Eq:ZeroTemp1MP2MP}
    \Delta_{\rm 1MP \leftrightarrow 2MP} = -3 \left( \mu + \frac{\mu_0 Q^{2}}{8 \pi a} \right) 
\end{equation}
(where Vac = vacuum and MP = monopole).
Since $\Delta$ is strictly positive, we can immediately conclude that no phase transitions are possible when $|\mu|<\frac{\mu_0 Q^{2}}{8 \pi a}$, in agreement with the results of Section \ref{density} and the arguments in Section \ref{breakdown}. We further see that the single monopole crystal phase only has a small window of stability when $|\mu|$ approaches $\frac{\mu_0 Q^{2}}{8 \pi a}$ from above. We finally note that the phase boundaries in Equations \ref{Eq:ZeroTempVac1MP} and \ref{Eq:ZeroTemp1MP2MP} again show remarkable agreement with those obtained by Raban \emph{et al.} in their model, with the exception that in their boundaries the $\frac{\mu_0 Q^{2}}{8 \pi a}$ is multiplied by the Madelung constant for their system, $\alpha=1.638$. Within Debye-H\"uckel theory we can argue that the effective Madelung constant is $\alpha = 1$, i.e. if the Madelung energy per site is written as  $E = -\alpha z^2 \mu_0 Q^2/8 \pi a$  (where $z$ is the charge number) then $\alpha =1$ equates to the energy per charge of a pair of equal and opposite charges. So we may conclude that the monopole density-jump transitions of the magnetolyte model of spin ice is the ``attempt'' of Debye-H\"uckel theory to capture the monopole crystallisation seen in staggered potential models. 

The Debye-H\"uckel model additionally gives insight into the nature of monopole crystallisation in staggered potential models. Naively, one might expect that the transition to a staggered-ordered state is induced solely by the change of symmetry introduced by the staggered potential $\Delta$. However, in addition to the symmetry change, there is already an instability to monopole crystallisation that results from the runaway excitation of monopoles due to charge-screening lowering the effective chemical potential of each additional monopole. In the Debye-H\"uckel ``dressed chemical potentials'' picture presented here, this process is the only mechanism available, since the symmetry of the staggered potential has not been taken into account.
Hence, in contrast to the case of the staggered-potential model \cite{Raban2019}, monopole crystallisation does not require a potential that explicitly breaks the symmetry of the dipolar spin ice model. The remarkable similarity between the phase boundaries of the staggered-potential and the dressed chemical potential models implies that reducing the effective chemical potential is what in fact drives the transition and the only role left for the staggered potential is to pick one of the two ``all-in, all-out'' states.

\section{Conclusions}

In conclusion, motivated by the arguments of Ryzhkin \emph{et al.} \cite{Ryzhkin2012}, we have have examined the possibility of density-jump transitions in the the Debye-H\"uckel theory of spin ice, as formulated by Kaiser \emph{et al}.~\cite{Kaiser2018}. While we conclude with certainty that such transitions do occur in the theory, it seems very likely that they represent the ``closest approach'' of Debye-H\"uckel theory to the true situation, as exposed by Raban \emph{et al.}~\cite{Raban2019}, where the transitions become monopole crystallisations. We have shown that Debye-H\"uckel theory and its extensions do a surprisingly good job at capturing the true behaviour if one equates full site occupation to crystallisation. 

Interpreted in this way, our analysis has value in understanding the monopole crystallisation transitions observed in the staggered-potential models of Ref. \cite{Raban2019}, showing that the transitions are largely a consequence of the Coulomb interaction and magnitude of the chemical potential, rather than of the explicit symmetry-breaking introduced by the staggered potential. Debye-H\"uckel theory may therefore serve as a useful bridge between staggered potential models and monopole-conserving models \cite{Borzi2013,Guruciaga2014} in which charge ordering occurs spontaneously at certain fixed monopole densities. Recent progress in experimentally determining the monopole density in HIO via magnetoresistance measurements \cite{pearce2022magnetic}, means that it should be possible to directly compare the behaviour of the monopole density in a staggered potential system with predictions of the Debye-H\"uckel model with dressed chemical potentials.

The alternative interpretation is that, either in spin ice systems or equivalent electrolytes, the transitions hint at true liquid-liquid transitions (i.e. jumps in charge density without crystallisation), as originally envisaged by Ryzhkin \emph{et al.}~\cite{Ryzhkin2012} For this to occur in real systems, one would presumably need to frustrate the crystallisation through supercooling \cite{Kassner2015} or by removing the lattice as in electrolyte solutions (where the dielectric permittivity provides a further degree of freedom~\cite{kozlov1990phase}) and glassy solid-state electrolytes \cite{Tomozawa1980,Deshpande2009,Grady2020}. Therefore, one place to search for such transitions might be the silicate glass system~\cite{Ingram1980} discussed in Ref. \cite{Kaiser2013} and Section \ref{electro}, where the parameters seem to be in the correct range.

An essential problem remains that the strong nonlinearities of Debye-H\"uckel theory, that give rise to the transitions, are also associated with the breakdown of the theory as an accurate description of the Coulomb fluid. Therefore while the theory presents an interesting example of a self-consistent model of liquid-liquid transitions that qualitatively describes transitions in certain physical systems, it is unlikely to be quantitative in this regard.

Since Debye-H\"{uckel} theory and ideas from the spin-fragmentation picture have been successfully applied to artificial spin ice systems~\cite{Farhan2019, Canals2016}, the inherent versatility of artificial spin systems could provide another avenue for exploring density-jump transitions in a regime where crystallisation is frustrated. This could take the form of a system that ``locally'' resembles spin ice (i.e., vertices of coordination four that obey ice rules in the ground state) but is spatially disordered globally so that monopoles cannot form a lattice. Such systems have been realised previously by designing systems in which the spin-lattice itself is non-periodic \cite{Dong2018, Saccone2019a, Saccone2019b} or by employing lattices where topological constraints prevent global ordering \cite{Morrison2013}.

While careful attention should be paid to how the Debye-H\"uckel model breaks at high monopole densities, the model may thus prove useful nevertheless in describing and interpreting density transitions in certain electrolytes, artificial spin systems and spin ice iridates.

\acknowledgements{We acknowledge support from the Leverhulme Trust through Grant No. RPG-2016-391. D.M.A. acknowledges funding through the NAME Programme Grant (EP/V001914/1).}

\bibliography{DebyeHuckel.bib}

\providecommand{\noopsort}[1]{}\providecommand{\singleletter}[1]{#1}%
\begin{thebibliography}{45}%
\makeatletter
\providecommand \@ifxundefined [1]{%
 \@ifx{#1\undefined}
}%
\providecommand \@ifnum [1]{%
 \ifnum #1\expandafter \@firstoftwo
 \else \expandafter \@secondoftwo
 \fi
}%
\providecommand \@ifx [1]{%
 \ifx #1\expandafter \@firstoftwo
 \else \expandafter \@secondoftwo
 \fi
}%
\providecommand \natexlab [1]{#1}%
\providecommand \enquote  [1]{``#1''}%
\providecommand \bibnamefont  [1]{#1}%
\providecommand \bibfnamefont [1]{#1}%
\providecommand \citenamefont [1]{#1}%
\providecommand \href@noop [0]{\@secondoftwo}%
\providecommand \href [0]{\begingroup \@sanitize@url \@href}%
\providecommand \@href[1]{\@@startlink{#1}\@@href}%
\providecommand \@@href[1]{\endgroup#1\@@endlink}%
\providecommand \@sanitize@url [0]{\catcode `\\12\catcode `\$12\catcode `\&12\catcode `\#12\catcode `\^12\catcode `\_12\catcode `\%12\relax}%
\providecommand \@@startlink[1]{}%
\providecommand \@@endlink[0]{}%
\providecommand \url  [0]{\begingroup\@sanitize@url \@url }%
\providecommand \@url [1]{\endgroup\@href {#1}{\urlprefix }}%
\providecommand \urlprefix  [0]{URL }%
\providecommand \Eprint [0]{\href }%
\providecommand \doibase [0]{http://dx.doi.org/}%
\providecommand \selectlanguage [0]{\@gobble}%
\providecommand \bibinfo  [0]{\@secondoftwo}%
\providecommand \bibfield  [0]{\@secondoftwo}%
\providecommand \translation [1]{[#1]}%
\providecommand \BibitemOpen [0]{}%
\providecommand \bibitemStop [0]{}%
\providecommand \bibitemNoStop [0]{.\EOS\space}%
\providecommand \EOS [0]{\spacefactor3000\relax}%
\providecommand \BibitemShut  [1]{\csname bibitem#1\endcsname}%
\let\auto@bib@innerbib\@empty
%</preamble>
\bibitem [{\citenamefont {Ramirez}(1994)}]{Ramirez1994}%
  \BibitemOpen
  \bibfield  {author} {\bibinfo {author} {\bibfnamefont {A.~P.}\ \bibnamefont {Ramirez}},\ }\href {\doibase 10.1146/annurev.ms.24.080194.002321} {\bibfield  {journal} {\bibinfo  {journal} {Annual Review of Materials Science}\ }\textbf {\bibinfo {volume} {24}},\ \bibinfo {pages} {453} (\bibinfo {year} {1994})}\BibitemShut {NoStop}%
\bibitem [{\citenamefont {Castelnovo}\ \emph {et~al.}(2012)\citenamefont {Castelnovo}, \citenamefont {Moessner},\ and\ \citenamefont {Sondhi}}]{CMSRev}%
  \BibitemOpen
  \bibfield  {author} {\bibinfo {author} {\bibfnamefont {C.}~\bibnamefont {Castelnovo}}, \bibinfo {author} {\bibfnamefont {R.}~\bibnamefont {Moessner}}, \ and\ \bibinfo {author} {\bibfnamefont {S.}~\bibnamefont {Sondhi}},\ }\href {\doibase 10.1146/annurev-conmatphys-020911-125058} {\bibfield  {journal} {\bibinfo  {journal} {Annual Review of Condensed Matter Physics}\ }\textbf {\bibinfo {volume} {3}},\ \bibinfo {pages} {35} (\bibinfo {year} {2012})}\BibitemShut {NoStop}%
\bibitem [{\citenamefont {Bramwell}\ and\ \citenamefont {Harris}(2020)}]{Bramwell2020}%
  \BibitemOpen
  \bibfield  {author} {\bibinfo {author} {\bibfnamefont {S.~T.}\ \bibnamefont {Bramwell}}\ and\ \bibinfo {author} {\bibfnamefont {M.~J.}\ \bibnamefont {Harris}},\ }\href {\doibase 10.1088/1361-648X/ab8423} {\bibfield  {journal} {\bibinfo  {journal} {J. Phys.: Condens. Matter}\ }\textbf {\bibinfo {volume} {32}},\ \bibinfo {pages} {374010} (\bibinfo {year} {2020})}\BibitemShut {NoStop}%
\bibitem [{\citenamefont {Pauling}(1935)}]{Pauling1935}%
  \BibitemOpen
  \bibfield  {author} {\bibinfo {author} {\bibfnamefont {L.}~\bibnamefont {Pauling}},\ }\href {\doibase 10.1021/ja01315a102} {\bibfield  {journal} {\bibinfo  {journal} {J. Am. Chem. Soc.}\ }\textbf {\bibinfo {volume} {57}},\ \bibinfo {pages} {2680} (\bibinfo {year} {1935})}\BibitemShut {NoStop}%
\bibitem [{\citenamefont {Baxter}(1982)}]{Baxter}%
  \BibitemOpen
  \bibfield  {author} {\bibinfo {author} {\bibfnamefont {R.~J.}\ \bibnamefont {Baxter}},\ }\href@noop {} {\emph {\bibinfo {title} {Exactly Solved Models in Statistical Mechanics}}}\ (\bibinfo  {publisher} {Academic Press, New York},\ \bibinfo {year} {1982})\BibitemShut {NoStop}%
\bibitem [{\citenamefont {Lieb}(1967)}]{Lieb1967}%
  \BibitemOpen
  \bibfield  {author} {\bibinfo {author} {\bibfnamefont {E.~H.}\ \bibnamefont {Lieb}},\ }\href {\doibase 10.1103/PhysRev.162.162} {\bibfield  {journal} {\bibinfo  {journal} {Phys. Rev.}\ }\textbf {\bibinfo {volume} {162}},\ \bibinfo {pages} {162} (\bibinfo {year} {1967})}\BibitemShut {NoStop}%
\bibitem [{\citenamefont {Henley}(2010)}]{Henley2010}%
  \BibitemOpen
  \bibfield  {author} {\bibinfo {author} {\bibfnamefont {C.~L.}\ \bibnamefont {Henley}},\ }\href {\doibase 10.1146/annurev-conmatphys-070909-104138} {\bibfield  {journal} {\bibinfo  {journal} {Annual Review of Condensed Matter Physics}\ }\textbf {\bibinfo {volume} {1}},\ \bibinfo {pages} {179} (\bibinfo {year} {2010})}\BibitemShut {NoStop}%
\bibitem [{\citenamefont {Ryzhkin}(2005)}]{Ryzhkin2005}%
  \BibitemOpen
  \bibfield  {author} {\bibinfo {author} {\bibfnamefont {I.~A.}\ \bibnamefont {Ryzhkin}},\ }\href {\doibase 10.1134/1.2103216} {\bibfield  {journal} {\bibinfo  {journal} {J. Exp. Theor. Phys.}\ }\textbf {\bibinfo {volume} {101}},\ \bibinfo {pages} {481} (\bibinfo {year} {2005})}\BibitemShut {NoStop}%
\bibitem [{\citenamefont {Castelnovo}\ \emph {et~al.}(2008)\citenamefont {Castelnovo}, \citenamefont {Moessner},\ and\ \citenamefont {Sondhi}}]{CMS2008}%
  \BibitemOpen
  \bibfield  {author} {\bibinfo {author} {\bibfnamefont {C.}~\bibnamefont {Castelnovo}}, \bibinfo {author} {\bibfnamefont {R.}~\bibnamefont {Moessner}}, \ and\ \bibinfo {author} {\bibfnamefont {S.~L.}\ \bibnamefont {Sondhi}},\ }\href {\doibase 10.1038/nature06433} {\bibfield  {journal} {\bibinfo  {journal} {Nature}\ }\textbf {\bibinfo {volume} {451}},\ \bibinfo {pages} {42} (\bibinfo {year} {2008})}\BibitemShut {NoStop}%
\bibitem [{\citenamefont {Castelnovo}\ and\ \citenamefont {Holdsworth}(2021)}]{Castelnovo2021}%
  \BibitemOpen
  \bibfield  {author} {\bibinfo {author} {\bibfnamefont {C.}~\bibnamefont {Castelnovo}}\ and\ \bibinfo {author} {\bibfnamefont {P.~C.~W.}\ \bibnamefont {Holdsworth}},\ }\enquote {\bibinfo {title} {Modelling of classical spin ice: Coulomb gas description of thermodynamic and dynamic properties},}\ in\ \href {\doibase 10.1007/978-3-030-70860-3_7} {\emph {\bibinfo {booktitle} {Spin Ice}}},\ \bibinfo {editor} {edited by\ \bibinfo {editor} {\bibfnamefont {M.}~\bibnamefont {Udagawa}}\ and\ \bibinfo {editor} {\bibfnamefont {L.}~\bibnamefont {Jaubert}}}\ (\bibinfo  {publisher} {Springer International Publishing},\ \bibinfo {address} {Cham},\ \bibinfo {year} {2021})\ pp.\ \bibinfo {pages} {143--188}\BibitemShut {NoStop}%
\bibitem [{\citenamefont {Castelnovo}\ \emph {et~al.}(2011)\citenamefont {Castelnovo}, \citenamefont {Moessner},\ and\ \citenamefont {Sondhi}}]{CMS2011}%
  \BibitemOpen
  \bibfield  {author} {\bibinfo {author} {\bibfnamefont {C.}~\bibnamefont {Castelnovo}}, \bibinfo {author} {\bibfnamefont {R.}~\bibnamefont {Moessner}}, \ and\ \bibinfo {author} {\bibfnamefont {S.~L.}\ \bibnamefont {Sondhi}},\ }\href {\doibase 10.1103/PhysRevB.84.144435} {\bibfield  {journal} {\bibinfo  {journal} {Phys. Rev. B}\ }\textbf {\bibinfo {volume} {84}},\ \bibinfo {pages} {144435} (\bibinfo {year} {2011})}\BibitemShut {NoStop}%
\bibitem [{\citenamefont {Debye}\ and\ \citenamefont {H\"{u}ckel}(1923)}]{DebyeHuckel1923}%
  \BibitemOpen
  \bibfield  {author} {\bibinfo {author} {\bibfnamefont {P.}~\bibnamefont {Debye}}\ and\ \bibinfo {author} {\bibfnamefont {E.}~\bibnamefont {H\"{u}ckel}},\ }\href@noop {} {\bibfield  {journal} {\bibinfo  {journal} {Physikalische Zeitschrift}\ }\textbf {\bibinfo {volume} {24}},\ \bibinfo {pages} {185} (\bibinfo {year} {1923})}\BibitemShut {NoStop}%
\bibitem [{\citenamefont {Moore}(1964)}]{Moore1964}%
  \BibitemOpen
  \bibfield  {author} {\bibinfo {author} {\bibfnamefont {W.}~\bibnamefont {Moore}},\ }\href@noop {} {\emph {\bibinfo {title} {Physical Chemistry}}},\ Prentice-Hall Chemistry Series\ (\bibinfo  {publisher} {Prentice-Hall},\ \bibinfo {year} {1964})\BibitemShut {NoStop}%
\bibitem [{\citenamefont {Kaiser}\ \emph {et~al.}(2018)\citenamefont {Kaiser}, \citenamefont {Bloxsom}, \citenamefont {Bovo}, \citenamefont {Bramwell}, \citenamefont {Holdsworth},\ and\ \citenamefont {Moessner}}]{Kaiser2018}%
  \BibitemOpen
  \bibfield  {author} {\bibinfo {author} {\bibfnamefont {V.}~\bibnamefont {Kaiser}}, \bibinfo {author} {\bibfnamefont {J.}~\bibnamefont {Bloxsom}}, \bibinfo {author} {\bibfnamefont {L.}~\bibnamefont {Bovo}}, \bibinfo {author} {\bibfnamefont {S.~T.}\ \bibnamefont {Bramwell}}, \bibinfo {author} {\bibfnamefont {P.~C.~W.}\ \bibnamefont {Holdsworth}}, \ and\ \bibinfo {author} {\bibfnamefont {R.}~\bibnamefont {Moessner}},\ }\href {\doibase 10.1103/PhysRevB.98.144413} {\bibfield  {journal} {\bibinfo  {journal} {Phys. Rev. B}\ }\textbf {\bibinfo {volume} {98}},\ \bibinfo {pages} {144413} (\bibinfo {year} {2018})}\BibitemShut {NoStop}%
\bibitem [{\citenamefont {Zhou}\ \emph {et~al.}(2011)\citenamefont {Zhou}, \citenamefont {Bramwell}, \citenamefont {Cheng}, \citenamefont {Wiebe}, \citenamefont {Li}, \citenamefont {Balicas}, \citenamefont {Bloxsom}, \citenamefont {Silverstein}, \citenamefont {Zhou}, \citenamefont {Goodenough},\ and\ \citenamefont {Gardner}}]{Zhou2011}%
  \BibitemOpen
  \bibfield  {author} {\bibinfo {author} {\bibfnamefont {H.~D.}\ \bibnamefont {Zhou}}, \bibinfo {author} {\bibfnamefont {S.~T.}\ \bibnamefont {Bramwell}}, \bibinfo {author} {\bibfnamefont {J.~G.}\ \bibnamefont {Cheng}}, \bibinfo {author} {\bibfnamefont {C.~R.}\ \bibnamefont {Wiebe}}, \bibinfo {author} {\bibfnamefont {G.}~\bibnamefont {Li}}, \bibinfo {author} {\bibfnamefont {L.}~\bibnamefont {Balicas}}, \bibinfo {author} {\bibfnamefont {J.~A.}\ \bibnamefont {Bloxsom}}, \bibinfo {author} {\bibfnamefont {H.~J.}\ \bibnamefont {Silverstein}}, \bibinfo {author} {\bibfnamefont {J.~S.}\ \bibnamefont {Zhou}}, \bibinfo {author} {\bibfnamefont {J.~B.}\ \bibnamefont {Goodenough}}, \ and\ \bibinfo {author} {\bibfnamefont {J.~S.}\ \bibnamefont {Gardner}},\ }\href {\doibase 10.1038/ncomms1483} {\bibfield  {journal} {\bibinfo  {journal} {Nat. Commun.}\ }\textbf {\bibinfo {volume} {2}},\ \bibinfo {pages} {478} (\bibinfo {year} {2011})}\BibitemShut {NoStop}%
\bibitem [{\citenamefont {Kirschner}\ \emph {et~al.}(2018)\citenamefont {Kirschner}, \citenamefont {Flicker}, \citenamefont {Yacoby}, \citenamefont {Yao},\ and\ \citenamefont {Blundell}}]{Kirschner2018}%
  \BibitemOpen
  \bibfield  {author} {\bibinfo {author} {\bibfnamefont {F.~K.~K.}\ \bibnamefont {Kirschner}}, \bibinfo {author} {\bibfnamefont {F.}~\bibnamefont {Flicker}}, \bibinfo {author} {\bibfnamefont {A.}~\bibnamefont {Yacoby}}, \bibinfo {author} {\bibfnamefont {N.~Y.}\ \bibnamefont {Yao}}, \ and\ \bibinfo {author} {\bibfnamefont {S.~J.}\ \bibnamefont {Blundell}},\ }\href {\doibase 10.1103/PhysRevB.97.140402} {\bibfield  {journal} {\bibinfo  {journal} {Phys. Rev. B}\ }\textbf {\bibinfo {volume} {97}},\ \bibinfo {pages} {140402} (\bibinfo {year} {2018})}\BibitemShut {NoStop}%
\bibitem [{\citenamefont {Farhan}\ \emph {et~al.}(2019)\citenamefont {Farhan}, \citenamefont {Saccone}, \citenamefont {Petersen}, \citenamefont {Dhuey}, \citenamefont {Chopdekar}, \citenamefont {Huang}, \citenamefont {Kent}, \citenamefont {Chen}, \citenamefont {Alava}, \citenamefont {Lippert}, \citenamefont {Scholl},\ and\ \citenamefont {van Dijken}}]{Farhan2019}%
  \BibitemOpen
  \bibfield  {author} {\bibinfo {author} {\bibfnamefont {A.}~\bibnamefont {Farhan}}, \bibinfo {author} {\bibfnamefont {M.}~\bibnamefont {Saccone}}, \bibinfo {author} {\bibfnamefont {C.~F.}\ \bibnamefont {Petersen}}, \bibinfo {author} {\bibfnamefont {S.}~\bibnamefont {Dhuey}}, \bibinfo {author} {\bibfnamefont {R.~V.}\ \bibnamefont {Chopdekar}}, \bibinfo {author} {\bibfnamefont {Y.-L.}\ \bibnamefont {Huang}}, \bibinfo {author} {\bibfnamefont {N.}~\bibnamefont {Kent}}, \bibinfo {author} {\bibfnamefont {Z.}~\bibnamefont {Chen}}, \bibinfo {author} {\bibfnamefont {M.~J.}\ \bibnamefont {Alava}}, \bibinfo {author} {\bibfnamefont {T.}~\bibnamefont {Lippert}}, \bibinfo {author} {\bibfnamefont {A.}~\bibnamefont {Scholl}}, \ and\ \bibinfo {author} {\bibfnamefont {S.}~\bibnamefont {van Dijken}},\ }\href {\doibase 10.1126/sciadv.aav6380} {\bibfield  {journal} {\bibinfo  {journal} {Science Advances}\ }\textbf {\bibinfo {volume} {5}} (\bibinfo {year} {2019}),\ 10.1126/sciadv.aav6380}\BibitemShut {NoStop}%
\bibitem [{\citenamefont {Ryzhkin}\ \emph {et~al.}(2012)\citenamefont {Ryzhkin}, \citenamefont {Klyuev}, \citenamefont {Ryzhkin},\ and\ \citenamefont {Tsybulin}}]{Ryzhkin2012}%
  \BibitemOpen
  \bibfield  {author} {\bibinfo {author} {\bibfnamefont {I.~A.}\ \bibnamefont {Ryzhkin}}, \bibinfo {author} {\bibfnamefont {A.~V.}\ \bibnamefont {Klyuev}}, \bibinfo {author} {\bibfnamefont {M.~I.}\ \bibnamefont {Ryzhkin}}, \ and\ \bibinfo {author} {\bibfnamefont {I.~V.}\ \bibnamefont {Tsybulin}},\ }\href {\doibase 10.1134/S0021364012060082} {\bibfield  {journal} {\bibinfo  {journal} {JETP Lett.}\ }\textbf {\bibinfo {volume} {95}},\ \bibinfo {pages} {302} (\bibinfo {year} {2012})}\BibitemShut {NoStop}%
\bibitem [{\citenamefont {Kozlov}\ \emph {et~al.}(1990)\citenamefont {Kozlov}, \citenamefont {Sokolova},\ and\ \citenamefont {Trufanov}}]{kozlov1990phase}%
  \BibitemOpen
  \bibfield  {author} {\bibinfo {author} {\bibfnamefont {V.}~\bibnamefont {Kozlov}}, \bibinfo {author} {\bibfnamefont {S.}~\bibnamefont {Sokolova}}, \ and\ \bibinfo {author} {\bibfnamefont {N.}~\bibnamefont {Trufanov}},\ }\href@noop {} {\bibfield  {journal} {\bibinfo  {journal} {Sov. Phys. JETP}\ }\textbf {\bibinfo {volume} {71}},\ \bibinfo {pages} {1224} (\bibinfo {year} {1990})}\BibitemShut {NoStop}%
\bibitem [{\citenamefont {Harris}\ \emph {et~al.}(1997)\citenamefont {Harris}, \citenamefont {Bramwell}, \citenamefont {McMorrow}, \citenamefont {Zeiske},\ and\ \citenamefont {Godfrey}}]{Harris1997}%
  \BibitemOpen
  \bibfield  {author} {\bibinfo {author} {\bibfnamefont {M.~J.}\ \bibnamefont {Harris}}, \bibinfo {author} {\bibfnamefont {S.~T.}\ \bibnamefont {Bramwell}}, \bibinfo {author} {\bibfnamefont {D.~F.}\ \bibnamefont {McMorrow}}, \bibinfo {author} {\bibfnamefont {T.}~\bibnamefont {Zeiske}}, \ and\ \bibinfo {author} {\bibfnamefont {K.~W.}\ \bibnamefont {Godfrey}},\ }\href {\doibase 10.1103/PhysRevLett.79.2554} {\bibfield  {journal} {\bibinfo  {journal} {Phys. Rev. Lett.}\ }\textbf {\bibinfo {volume} {79}},\ \bibinfo {pages} {2554} (\bibinfo {year} {1997})}\BibitemShut {NoStop}%
\bibitem [{\citenamefont {Matsuhira}\ \emph {et~al.}(2000)\citenamefont {Matsuhira}, \citenamefont {Hinatsu}, \citenamefont {Tenya},\ and\ \citenamefont {Sakakibara}}]{Matsuhira2000}%
  \BibitemOpen
  \bibfield  {author} {\bibinfo {author} {\bibfnamefont {K.}~\bibnamefont {Matsuhira}}, \bibinfo {author} {\bibfnamefont {Y.}~\bibnamefont {Hinatsu}}, \bibinfo {author} {\bibfnamefont {K.}~\bibnamefont {Tenya}}, \ and\ \bibinfo {author} {\bibfnamefont {T.}~\bibnamefont {Sakakibara}},\ }\href {\doibase 10.1088/0953-8984/12/40/103} {\bibfield  {journal} {\bibinfo  {journal} {J. Phys.: Condens. Matter}\ }\textbf {\bibinfo {volume} {12}},\ \bibinfo {pages} {L649} (\bibinfo {year} {2000})}\BibitemShut {NoStop}%
\bibitem [{\citenamefont {Clancy}\ \emph {et~al.}(2009)\citenamefont {Clancy}, \citenamefont {Ruff}, \citenamefont {Dunsiger}, \citenamefont {Zhao}, \citenamefont {Dabkowska}, \citenamefont {Gardner}, \citenamefont {Qiu}, \citenamefont {Copley}, \citenamefont {Jenkins},\ and\ \citenamefont {Gaulin}}]{Clancy2009}%
  \BibitemOpen
  \bibfield  {author} {\bibinfo {author} {\bibfnamefont {J.~P.}\ \bibnamefont {Clancy}}, \bibinfo {author} {\bibfnamefont {J.~P.~C.}\ \bibnamefont {Ruff}}, \bibinfo {author} {\bibfnamefont {S.~R.}\ \bibnamefont {Dunsiger}}, \bibinfo {author} {\bibfnamefont {Y.}~\bibnamefont {Zhao}}, \bibinfo {author} {\bibfnamefont {H.~A.}\ \bibnamefont {Dabkowska}}, \bibinfo {author} {\bibfnamefont {J.~S.}\ \bibnamefont {Gardner}}, \bibinfo {author} {\bibfnamefont {Y.}~\bibnamefont {Qiu}}, \bibinfo {author} {\bibfnamefont {J.~R.~D.}\ \bibnamefont {Copley}}, \bibinfo {author} {\bibfnamefont {T.}~\bibnamefont {Jenkins}}, \ and\ \bibinfo {author} {\bibfnamefont {B.~D.}\ \bibnamefont {Gaulin}},\ }\href {\doibase 10.1103/PhysRevB.79.014408} {\bibfield  {journal} {\bibinfo  {journal} {Phys. Rev. B}\ }\textbf {\bibinfo {volume} {79}},\ \bibinfo {pages} {014408} (\bibinfo {year} {2009})}\BibitemShut {NoStop}%
\bibitem [{\citenamefont {Paulsen}\ \emph {et~al.}(2019)\citenamefont {Paulsen}, \citenamefont {Giblin}, \citenamefont {Lhotel}, \citenamefont {Prabhakaran}, \citenamefont {Matsuhira}, \citenamefont {Balakrishnan},\ and\ \citenamefont {Bramwell}}]{Paulsen2019}%
  \BibitemOpen
  \bibfield  {author} {\bibinfo {author} {\bibfnamefont {C.}~\bibnamefont {Paulsen}}, \bibinfo {author} {\bibfnamefont {S.~R.}\ \bibnamefont {Giblin}}, \bibinfo {author} {\bibfnamefont {E.}~\bibnamefont {Lhotel}}, \bibinfo {author} {\bibfnamefont {D.}~\bibnamefont {Prabhakaran}}, \bibinfo {author} {\bibfnamefont {K.}~\bibnamefont {Matsuhira}}, \bibinfo {author} {\bibfnamefont {G.}~\bibnamefont {Balakrishnan}}, \ and\ \bibinfo {author} {\bibfnamefont {S.~T.}\ \bibnamefont {Bramwell}},\ }\href {\doibase 10.1038/s41467-019-09323-6} {\bibfield  {journal} {\bibinfo  {journal} {Nature Communications}\ }\textbf {\bibinfo {volume} {10}},\ \bibinfo {pages} {1509} (\bibinfo {year} {2019})}\BibitemShut {NoStop}%
\bibitem [{\citenamefont {Brooks-Bartlett}\ \emph {et~al.}(2014)\citenamefont {Brooks-Bartlett}, \citenamefont {Banks}, \citenamefont {Jaubert}, \citenamefont {Harman-Clarke},\ and\ \citenamefont {Holdsworth}}]{BrooksBartlett2014}%
  \BibitemOpen
  \bibfield  {author} {\bibinfo {author} {\bibfnamefont {M.~E.}\ \bibnamefont {Brooks-Bartlett}}, \bibinfo {author} {\bibfnamefont {S.~T.}\ \bibnamefont {Banks}}, \bibinfo {author} {\bibfnamefont {L.~D.~C.}\ \bibnamefont {Jaubert}}, \bibinfo {author} {\bibfnamefont {A.}~\bibnamefont {Harman-Clarke}}, \ and\ \bibinfo {author} {\bibfnamefont {P.~C.~W.}\ \bibnamefont {Holdsworth}},\ }\href {\doibase 10.1103/PhysRevX.4.011007} {\bibfield  {journal} {\bibinfo  {journal} {Phys. Rev. X}\ }\textbf {\bibinfo {volume} {4}},\ \bibinfo {pages} {011007} (\bibinfo {year} {2014})}\BibitemShut {NoStop}%
\bibitem [{\citenamefont {Raban}\ \emph {et~al.}(2019)\citenamefont {Raban}, \citenamefont {Suen}, \citenamefont {Berthier},\ and\ \citenamefont {Holdsworth}}]{Raban2019}%
  \BibitemOpen
  \bibfield  {author} {\bibinfo {author} {\bibfnamefont {V.}~\bibnamefont {Raban}}, \bibinfo {author} {\bibfnamefont {C.~T.}\ \bibnamefont {Suen}}, \bibinfo {author} {\bibfnamefont {L.}~\bibnamefont {Berthier}}, \ and\ \bibinfo {author} {\bibfnamefont {P.~C.~W.}\ \bibnamefont {Holdsworth}},\ }\href {\doibase 10.1103/PhysRevB.99.224425} {\bibfield  {journal} {\bibinfo  {journal} {Phys. Rev. B}\ }\textbf {\bibinfo {volume} {99}},\ \bibinfo {pages} {224425} (\bibinfo {year} {2019})}\BibitemShut {NoStop}%
\bibitem [{\citenamefont {Borzi}\ \emph {et~al.}(2013)\citenamefont {Borzi}, \citenamefont {Slobinsky},\ and\ \citenamefont {Grigera}}]{Borzi2013}%
  \BibitemOpen
  \bibfield  {author} {\bibinfo {author} {\bibfnamefont {R.~A.}\ \bibnamefont {Borzi}}, \bibinfo {author} {\bibfnamefont {D.}~\bibnamefont {Slobinsky}}, \ and\ \bibinfo {author} {\bibfnamefont {S.~A.}\ \bibnamefont {Grigera}},\ }\href {\doibase 10.1103/PhysRevLett.111.147204} {\bibfield  {journal} {\bibinfo  {journal} {Phys. Rev. Lett.}\ }\textbf {\bibinfo {volume} {111}},\ \bibinfo {pages} {147204} (\bibinfo {year} {2013})}\BibitemShut {NoStop}%
\bibitem [{Note1()}]{Note1}%
  \BibitemOpen
  \bibinfo {note} {Note that this corrects an error in the expression given for the Coulomb energy in Equation 13 of Ref.~\cite {Kaiser2018}, which is too small by a factor of four; the error does not propagate into the results of that paper.}\BibitemShut {Stop}%
\bibitem [{\citenamefont {Kaiser}\ \emph {et~al.}(2013)\citenamefont {Kaiser}, \citenamefont {Bramwell}, \citenamefont {Holdsworth},\ and\ \citenamefont {Moessner}}]{Kaiser2013}%
  \BibitemOpen
  \bibfield  {author} {\bibinfo {author} {\bibfnamefont {V.}~\bibnamefont {Kaiser}}, \bibinfo {author} {\bibfnamefont {S.~T.}\ \bibnamefont {Bramwell}}, \bibinfo {author} {\bibfnamefont {P.~C.~W.}\ \bibnamefont {Holdsworth}}, \ and\ \bibinfo {author} {\bibfnamefont {R.}~\bibnamefont {Moessner}},\ }\href {\doibase 10.1038/nmat3729} {\bibfield  {journal} {\bibinfo  {journal} {Nature Mater.}\ }\textbf {\bibinfo {volume} {12}},\ \bibinfo {pages} {1033} (\bibinfo {year} {2013})}\BibitemShut {NoStop}%
\bibitem [{\citenamefont {Ingram}\ \emph {et~al.}(1980)\citenamefont {Ingram}, \citenamefont {Moynihan},\ and\ \citenamefont {Lesikar}}]{Ingram1980}%
  \BibitemOpen
  \bibfield  {author} {\bibinfo {author} {\bibfnamefont {M.}~\bibnamefont {Ingram}}, \bibinfo {author} {\bibfnamefont {C.}~\bibnamefont {Moynihan}}, \ and\ \bibinfo {author} {\bibfnamefont {A.}~\bibnamefont {Lesikar}},\ }\href {\doibase https://doi.org/10.1016/0022-3093(80)90447-0} {\bibfield  {journal} {\bibinfo  {journal} {Journal of Non-Crystalline Solids}\ }\textbf {\bibinfo {volume} {38-39}},\ \bibinfo {pages} {371} (\bibinfo {year} {1980})},\ \bibinfo {note} {xIIth International Congress on Glass}\BibitemShut {NoStop}%
\bibitem [{\citenamefont {Guruciaga}\ \emph {et~al.}(2014)\citenamefont {Guruciaga}, \citenamefont {Grigera},\ and\ \citenamefont {Borzi}}]{Guruciaga2014}%
  \BibitemOpen
  \bibfield  {author} {\bibinfo {author} {\bibfnamefont {P.~C.}\ \bibnamefont {Guruciaga}}, \bibinfo {author} {\bibfnamefont {S.~A.}\ \bibnamefont {Grigera}}, \ and\ \bibinfo {author} {\bibfnamefont {R.~A.}\ \bibnamefont {Borzi}},\ }\href {\doibase 10.1103/PhysRevB.90.184423} {\bibfield  {journal} {\bibinfo  {journal} {Phys. Rev. B}\ }\textbf {\bibinfo {volume} {90}},\ \bibinfo {pages} {184423} (\bibinfo {year} {2014})}\BibitemShut {NoStop}%
\bibitem [{\citenamefont {Petit}\ \emph {et~al.}(2016)\citenamefont {Petit}, \citenamefont {Lhotel}, \citenamefont {Canals}, \citenamefont {Hatnean}, \citenamefont {Ollivier}, \citenamefont {Mutka}, \citenamefont {Ressouche}, \citenamefont {Wildes}, \citenamefont {Lees},\ and\ \citenamefont {Balakrishnan}}]{Petit2016}%
  \BibitemOpen
  \bibfield  {author} {\bibinfo {author} {\bibfnamefont {S.}~\bibnamefont {Petit}}, \bibinfo {author} {\bibfnamefont {E.}~\bibnamefont {Lhotel}}, \bibinfo {author} {\bibfnamefont {B.}~\bibnamefont {Canals}}, \bibinfo {author} {\bibfnamefont {M.~C.}\ \bibnamefont {Hatnean}}, \bibinfo {author} {\bibfnamefont {J.}~\bibnamefont {Ollivier}}, \bibinfo {author} {\bibfnamefont {H.}~\bibnamefont {Mutka}}, \bibinfo {author} {\bibfnamefont {E.}~\bibnamefont {Ressouche}}, \bibinfo {author} {\bibfnamefont {A.~R.}\ \bibnamefont {Wildes}}, \bibinfo {author} {\bibfnamefont {M.~R.}\ \bibnamefont {Lees}}, \ and\ \bibinfo {author} {\bibfnamefont {G.}~\bibnamefont {Balakrishnan}},\ }\href {\doibase 10.1038/nphys3710} {\bibfield  {journal} {\bibinfo  {journal} {Nature Physics}\ }\textbf {\bibinfo {volume} {12}},\ \bibinfo {pages} {746} (\bibinfo {year} {2016})}\BibitemShut {NoStop}%
\bibitem [{\citenamefont {Lhotel}\ \emph {et~al.}(2020)\citenamefont {Lhotel}, \citenamefont {Jaubert},\ and\ \citenamefont {Holdsworth}}]{Lhotel2020}%
  \BibitemOpen
  \bibfield  {author} {\bibinfo {author} {\bibfnamefont {E.}~\bibnamefont {Lhotel}}, \bibinfo {author} {\bibfnamefont {L.~D.~C.}\ \bibnamefont {Jaubert}}, \ and\ \bibinfo {author} {\bibfnamefont {P.~C.~W.}\ \bibnamefont {Holdsworth}},\ }\href {\doibase 10.1007/s10909-020-02521-3} {\bibfield  {journal} {\bibinfo  {journal} {J. Low Temp. Phys.}\ }\textbf {\bibinfo {volume} {201}},\ \bibinfo {pages} {710} (\bibinfo {year} {2020})}\BibitemShut {NoStop}%
\bibitem [{\citenamefont {Raban}\ \emph {et~al.}(2022)\citenamefont {Raban}, \citenamefont {Berthier},\ and\ \citenamefont {Holdsworth}}]{raban2022violation}%
  \BibitemOpen
  \bibfield  {author} {\bibinfo {author} {\bibfnamefont {V.}~\bibnamefont {Raban}}, \bibinfo {author} {\bibfnamefont {L.}~\bibnamefont {Berthier}}, \ and\ \bibinfo {author} {\bibfnamefont {P.~C.}\ \bibnamefont {Holdsworth}},\ }\href@noop {} {\bibfield  {journal} {\bibinfo  {journal} {Physical Review B}\ }\textbf {\bibinfo {volume} {105}},\ \bibinfo {pages} {134431} (\bibinfo {year} {2022})}\BibitemShut {NoStop}%
\bibitem [{\citenamefont {Lefrançois}\ \emph {et~al.}(2017)\citenamefont {Lefrançois}, \citenamefont {Cathelin}, \citenamefont {Lhotel}, \citenamefont {Robert}, \citenamefont {Lejay}, \citenamefont {Colin}, \citenamefont {Canals}, \citenamefont {Damay}, \citenamefont {Ollivier}, \citenamefont {Fåk}, \citenamefont {Chapon}, \citenamefont {Ballou},\ and\ \citenamefont {Simonet}}]{Lefrancois2017}%
  \BibitemOpen
  \bibfield  {author} {\bibinfo {author} {\bibfnamefont {E.}~\bibnamefont {Lefrançois}}, \bibinfo {author} {\bibfnamefont {V.}~\bibnamefont {Cathelin}}, \bibinfo {author} {\bibfnamefont {E.}~\bibnamefont {Lhotel}}, \bibinfo {author} {\bibfnamefont {J.}~\bibnamefont {Robert}}, \bibinfo {author} {\bibfnamefont {P.}~\bibnamefont {Lejay}}, \bibinfo {author} {\bibfnamefont {C.~V.}\ \bibnamefont {Colin}}, \bibinfo {author} {\bibfnamefont {B.}~\bibnamefont {Canals}}, \bibinfo {author} {\bibfnamefont {F.}~\bibnamefont {Damay}}, \bibinfo {author} {\bibfnamefont {J.}~\bibnamefont {Ollivier}}, \bibinfo {author} {\bibfnamefont {B.}~\bibnamefont {Fåk}}, \bibinfo {author} {\bibfnamefont {L.~C.}\ \bibnamefont {Chapon}}, \bibinfo {author} {\bibfnamefont {R.}~\bibnamefont {Ballou}}, \ and\ \bibinfo {author} {\bibfnamefont {V.}~\bibnamefont {Simonet}},\ }\href {\doibase 10.1038/s41467-017-00277-1} {\bibfield  {journal} {\bibinfo  {journal} {Nature Communications}\ }\textbf {\bibinfo {volume} {8}},\ \bibinfo {pages} {209}
  (\bibinfo {year} {2017})}\BibitemShut {NoStop}%
\bibitem [{\citenamefont {Raban}(2018)}]{RabanThesis}%
  \BibitemOpen
  \bibfield  {author} {\bibinfo {author} {\bibfnamefont {V.}~\bibnamefont {Raban}},\ }\emph {\bibinfo {title} {Dynamique hors \'equilibre des monop\^oles magn\'etiques dans la glace de spin}},\ \href {https://tel.archives-ouvertes.fr/tel-01974349} {Ph.D. thesis},\ \bibinfo  {school} {Universit\'e de Lyon} (\bibinfo {year} {2018}),\ \bibinfo {note} {{NNT}: 2018LY-SEN052. tel-01974349.}\BibitemShut {Stop}%
\bibitem [{\citenamefont {Pearce}\ \emph {et~al.}(2022)\citenamefont {Pearce}, \citenamefont {G{\"o}tze}, \citenamefont {Szab{\'o}}, \citenamefont {Sikkenk}, \citenamefont {Lees}, \citenamefont {Boothroyd}, \citenamefont {Prabhakaran}, \citenamefont {Castelnovo},\ and\ \citenamefont {Goddard}}]{pearce2022magnetic}%
  \BibitemOpen
  \bibfield  {author} {\bibinfo {author} {\bibfnamefont {M.~J.}\ \bibnamefont {Pearce}}, \bibinfo {author} {\bibfnamefont {K.}~\bibnamefont {G{\"o}tze}}, \bibinfo {author} {\bibfnamefont {A.}~\bibnamefont {Szab{\'o}}}, \bibinfo {author} {\bibfnamefont {T.~S.}\ \bibnamefont {Sikkenk}}, \bibinfo {author} {\bibfnamefont {M.~R.}\ \bibnamefont {Lees}}, \bibinfo {author} {\bibfnamefont {A.~T.}\ \bibnamefont {Boothroyd}}, \bibinfo {author} {\bibfnamefont {D.}~\bibnamefont {Prabhakaran}}, \bibinfo {author} {\bibfnamefont {C.}~\bibnamefont {Castelnovo}}, \ and\ \bibinfo {author} {\bibfnamefont {P.~A.}\ \bibnamefont {Goddard}},\ }\href@noop {} {\bibfield  {journal} {\bibinfo  {journal} {Nature Communications}\ }\textbf {\bibinfo {volume} {13}},\ \bibinfo {pages} {444} (\bibinfo {year} {2022})}\BibitemShut {NoStop}%
\bibitem [{\citenamefont {Kassner}\ \emph {et~al.}(2015)\citenamefont {Kassner}, \citenamefont {Eyvazov}, \citenamefont {Pichler}, \citenamefont {Munsie}, \citenamefont {Dabkowska}, \citenamefont {Luke},\ and\ \citenamefont {Davis}}]{Kassner2015}%
  \BibitemOpen
  \bibfield  {author} {\bibinfo {author} {\bibfnamefont {E.~R.}\ \bibnamefont {Kassner}}, \bibinfo {author} {\bibfnamefont {A.~B.}\ \bibnamefont {Eyvazov}}, \bibinfo {author} {\bibfnamefont {B.}~\bibnamefont {Pichler}}, \bibinfo {author} {\bibfnamefont {T.~J.~S.}\ \bibnamefont {Munsie}}, \bibinfo {author} {\bibfnamefont {H.~A.}\ \bibnamefont {Dabkowska}}, \bibinfo {author} {\bibfnamefont {G.~M.}\ \bibnamefont {Luke}}, \ and\ \bibinfo {author} {\bibfnamefont {J.~C.~S.}\ \bibnamefont {Davis}},\ }\href {\doibase 10.1073/pnas.1511006112} {\bibfield  {journal} {\bibinfo  {journal} {Proc. Natl. Acad. Sci. U.S.A.}\ }\textbf {\bibinfo {volume} {112}},\ \bibinfo {pages} {8549} (\bibinfo {year} {2015})}\BibitemShut {NoStop}%
\bibitem [{\citenamefont {Tomozawa}\ \emph {et~al.}(1980)\citenamefont {Tomozawa}, \citenamefont {Cordaro},\ and\ \citenamefont {Singh}}]{Tomozawa1980}%
  \BibitemOpen
  \bibfield  {author} {\bibinfo {author} {\bibfnamefont {M.}~\bibnamefont {Tomozawa}}, \bibinfo {author} {\bibfnamefont {J.}~\bibnamefont {Cordaro}}, \ and\ \bibinfo {author} {\bibfnamefont {M.}~\bibnamefont {Singh}},\ }\href {\doibase https://doi.org/10.1016/0022-3093(80)90102-7} {\bibfield  {journal} {\bibinfo  {journal} {Journal of Non-Crystalline Solids}\ }\textbf {\bibinfo {volume} {40}},\ \bibinfo {pages} {189} (\bibinfo {year} {1980})},\ \bibinfo {note} {proceedings of the Fifth University Conference on Glass Science}\BibitemShut {NoStop}%
\bibitem [{\citenamefont {Deshpande}(2009)}]{Deshpande2009}%
  \BibitemOpen
  \bibfield  {author} {\bibinfo {author} {\bibfnamefont {V.~K.}\ \bibnamefont {Deshpande}},\ }\href {\doibase 10.1088/1757-899x/2/1/012011} {\bibfield  {journal} {\bibinfo  {journal} {{IOP} Conference Series: Materials Science and Engineering}\ }\textbf {\bibinfo {volume} {2}},\ \bibinfo {pages} {012011} (\bibinfo {year} {2009})}\BibitemShut {NoStop}%
\bibitem [{\citenamefont {Grady}\ \emph {et~al.}(2020)\citenamefont {Grady}, \citenamefont {Wilkinson}, \citenamefont {Randall},\ and\ \citenamefont {Mauro}}]{Grady2020}%
  \BibitemOpen
  \bibfield  {author} {\bibinfo {author} {\bibfnamefont {Z.~A.}\ \bibnamefont {Grady}}, \bibinfo {author} {\bibfnamefont {C.~J.}\ \bibnamefont {Wilkinson}}, \bibinfo {author} {\bibfnamefont {C.~A.}\ \bibnamefont {Randall}}, \ and\ \bibinfo {author} {\bibfnamefont {J.~C.}\ \bibnamefont {Mauro}},\ }\href {\doibase 10.3389/fenrg.2020.00218} {\bibfield  {journal} {\bibinfo  {journal} {Frontiers in Energy Research}\ }\textbf {\bibinfo {volume} {8}},\ \bibinfo {pages} {218} (\bibinfo {year} {2020})}\BibitemShut {NoStop}%
\bibitem [{\citenamefont {Canals}\ \emph {et~al.}(2016)\citenamefont {Canals}, \citenamefont {Chioar}, \citenamefont {Nguyen}, \citenamefont {Hehn}, \citenamefont {Lacour}, \citenamefont {Montaigne}, \citenamefont {Locatelli}, \citenamefont {Menteş}, \citenamefont {Burgos},\ and\ \citenamefont {Rougemaille}}]{Canals2016}%
  \BibitemOpen
  \bibfield  {author} {\bibinfo {author} {\bibfnamefont {B.}~\bibnamefont {Canals}}, \bibinfo {author} {\bibfnamefont {I.-A.}\ \bibnamefont {Chioar}}, \bibinfo {author} {\bibfnamefont {V.}~\bibnamefont {Nguyen}}, \bibinfo {author} {\bibfnamefont {M.}~\bibnamefont {Hehn}}, \bibinfo {author} {\bibfnamefont {D.}~\bibnamefont {Lacour}}, \bibinfo {author} {\bibfnamefont {F.}~\bibnamefont {Montaigne}}, \bibinfo {author} {\bibfnamefont {A.}~\bibnamefont {Locatelli}}, \bibinfo {author} {\bibfnamefont {T.~O.}\ \bibnamefont {Menteş}}, \bibinfo {author} {\bibfnamefont {B.~S.}\ \bibnamefont {Burgos}}, \ and\ \bibinfo {author} {\bibfnamefont {N.}~\bibnamefont {Rougemaille}},\ }\href {\doibase 10.1038/ncomms11446} {\bibfield  {journal} {\bibinfo  {journal} {Nature Communications}\ }\textbf {\bibinfo {volume} {7}},\ \bibinfo {pages} {11446} (\bibinfo {year} {2016})}\BibitemShut {NoStop}%
\bibitem [{\citenamefont {Shi}\ \emph {et~al.}(2018)\citenamefont {Shi}, \citenamefont {Budrikis}, \citenamefont {Stein}, \citenamefont {Morley}, \citenamefont {Olmsted}, \citenamefont {Burnell},\ and\ \citenamefont {Marrows}}]{Dong2018}%
  \BibitemOpen
  \bibfield  {author} {\bibinfo {author} {\bibfnamefont {D.}~\bibnamefont {Shi}}, \bibinfo {author} {\bibfnamefont {Z.}~\bibnamefont {Budrikis}}, \bibinfo {author} {\bibfnamefont {A.}~\bibnamefont {Stein}}, \bibinfo {author} {\bibfnamefont {S.~A.}\ \bibnamefont {Morley}}, \bibinfo {author} {\bibfnamefont {P.~D.}\ \bibnamefont {Olmsted}}, \bibinfo {author} {\bibfnamefont {G.}~\bibnamefont {Burnell}}, \ and\ \bibinfo {author} {\bibfnamefont {C.~H.}\ \bibnamefont {Marrows}},\ }\href {\doibase 10.1038/s41567-017-0009-4} {\bibfield  {journal} {\bibinfo  {journal} {Nature Physics}\ }\textbf {\bibinfo {volume} {14}},\ \bibinfo {pages} {309} (\bibinfo {year} {2018})}\BibitemShut {NoStop}%
\bibitem [{\citenamefont {Saccone}\ \emph {et~al.}(2019)\citenamefont {Saccone}, \citenamefont {Scholl}, \citenamefont {Velten}, \citenamefont {Dhuey}, \citenamefont {Hofhuis}, \citenamefont {Wuth}, \citenamefont {Huang}, \citenamefont {Chen}, \citenamefont {Chopdekar},\ and\ \citenamefont {Farhan}}]{Saccone2019a}%
  \BibitemOpen
  \bibfield  {author} {\bibinfo {author} {\bibfnamefont {M.}~\bibnamefont {Saccone}}, \bibinfo {author} {\bibfnamefont {A.}~\bibnamefont {Scholl}}, \bibinfo {author} {\bibfnamefont {S.}~\bibnamefont {Velten}}, \bibinfo {author} {\bibfnamefont {S.}~\bibnamefont {Dhuey}}, \bibinfo {author} {\bibfnamefont {K.}~\bibnamefont {Hofhuis}}, \bibinfo {author} {\bibfnamefont {C.}~\bibnamefont {Wuth}}, \bibinfo {author} {\bibfnamefont {Y.-L.}\ \bibnamefont {Huang}}, \bibinfo {author} {\bibfnamefont {Z.}~\bibnamefont {Chen}}, \bibinfo {author} {\bibfnamefont {R.~V.}\ \bibnamefont {Chopdekar}}, \ and\ \bibinfo {author} {\bibfnamefont {A.}~\bibnamefont {Farhan}},\ }\href {\doibase 10.1103/PhysRevB.99.224403} {\bibfield  {journal} {\bibinfo  {journal} {Physical Review B}\ }\textbf {\bibinfo {volume} {99}},\ \bibinfo {pages} {224403} (\bibinfo {year} {2019})}\BibitemShut {NoStop}%
\bibitem [{\citenamefont {Saccone}\ \emph {et~al.}(2020)\citenamefont {Saccone}, \citenamefont {Hofhuis}, \citenamefont {Bracher}, \citenamefont {Kleibert}, \citenamefont {van Dijken},\ and\ \citenamefont {Farhan}}]{Saccone2019b}%
  \BibitemOpen
  \bibfield  {author} {\bibinfo {author} {\bibfnamefont {M.}~\bibnamefont {Saccone}}, \bibinfo {author} {\bibfnamefont {K.}~\bibnamefont {Hofhuis}}, \bibinfo {author} {\bibfnamefont {D.}~\bibnamefont {Bracher}}, \bibinfo {author} {\bibfnamefont {A.}~\bibnamefont {Kleibert}}, \bibinfo {author} {\bibfnamefont {S.}~\bibnamefont {van Dijken}}, \ and\ \bibinfo {author} {\bibfnamefont {A.}~\bibnamefont {Farhan}},\ }\href {\doibase 10.1039/C9NR07510K} {\bibfield  {journal} {\bibinfo  {journal} {Nanoscale}\ }\textbf {\bibinfo {volume} {12}},\ \bibinfo {pages} {189} (\bibinfo {year} {2020})}\BibitemShut {NoStop}%
\bibitem [{\citenamefont {Morrison}\ \emph {et~al.}(2013)\citenamefont {Morrison}, \citenamefont {Nelson},\ and\ \citenamefont {Nisoli}}]{Morrison2013}%
  \BibitemOpen
  \bibfield  {author} {\bibinfo {author} {\bibfnamefont {M.~J.}\ \bibnamefont {Morrison}}, \bibinfo {author} {\bibfnamefont {T.~R.}\ \bibnamefont {Nelson}}, \ and\ \bibinfo {author} {\bibfnamefont {C.}~\bibnamefont {Nisoli}},\ }\href {\doibase 10.1088/1367-2630/15/4/045009} {\bibfield  {journal} {\bibinfo  {journal} {New Journal of Physics}\ }\textbf {\bibinfo {volume} {15}},\ \bibinfo {pages} {045009} (\bibinfo {year} {2013})}\BibitemShut {NoStop}%
\end{thebibliography}%

\end{document}